\documentclass[12pt,preprint]{aastex}
\usepackage{mathptm}
\usepackage{epsfig}
\usepackage{rotating}
\slugcomment{\it Accepted by ApJ, 06/17/05.}
\begin{document}

\title{IRAS~16293$-$2422: proper motions, jet precession, the hot core,
and the unambiguous detection of infall}
\author{Claire J. Chandler}
\affil{National Radio Astronomy Observatory\footnote{The NRAO is operated
by Associated Universities Inc., under cooperative agreement with the
National Science Foundation}, PO Box O, Socorro, NM 87801}
\author{Crystal L. Brogan\footnote{JCMT Fellow at the IfA, Hilo, Hawaii.}}
\affil{University of Hawaii, Institute for Astronomy, 640 North
A'ohoku Place, Hilo, HI 96720}
\author{Yancy L. Shirley\footnote{Jansky Fellow at NRAO, Socorro,
New Mexico.}}
\affil{National Radio Astronomy Observatory, PO Box O, Socorro, NM
87801}
\and
\author{Laurent Loinard}
\affil{Centro de Radioastronom\'\i a y Astrof\'\i sica, Universidad
Nacional Aut\'onoma de M\'exico, \\ Apdo Postal 72--3 (Xangari), 58089
Morelia, Michoac\'an, M\'exico}
\affil{E-mail: cchandle@nrao.edu, cbrogan@ifa.hawaii.edu,
yshirley@nrao.edu, l.loinard@astrosmo.unam.mx}

\begin{abstract}

We present high spatial resolution observations of the multiple
protostellar system IRAS~16293$-$2422 using the Submillimeter Array
(SMA) at 300~GHz, and the Very Large Array (VLA) at frequencies from
1.5 to 43~GHz.  This source was already known to be a binary system
with its main components, A and B, separated by $\sim 5''$.  The new SMA
data now separate source A into two submillimeter continuum components,
which we denote Aa and Ab.  The strongest of these, Aa, peaks between
the centimeter radio sources A1 and A2, but the resolution of the
current submillimeter data is insufficient to distinguish whether this
is a separate source or the centroid of submillimeter dust emission
associated with A1 and A2.  Archival VLA data spanning 18 years show
proper motion of sources A and B of 17~mas~yr$^{-1}$, associated with the
motion of the $\rho$ Ophiuchi cloud.  We also find, however, significant
relative motion between the centimeter sources A1 and A2 which excludes
the possibility that these two sources are gravitationally bound unless
A1 is in a highly eccentric orbit and is observed at periastron, the
probability of which is low.  A2 remains stationary relative to source B,
and we identify it as the protostar which drives the large-scale NE--SW
CO outflow.  A1 is shock-ionized gas which traces the location of the
interaction between a precessing jet and nearby dense gas.  This jet
probably drives the large-scale E--W outflow, and indeed its motion
is consistent with the wide opening angle of this flow.  The origin
of this jet must be located close to A2, and may be the submillimeter
continuum source Aa.  Thus source A is now shown to comprise three
(proto)stellar components within $1''$.  Source B, on the other hand,
is single, exhibits optically-thick dust emission even at 8~GHz, has a
high luminosity, and yet shows no sign of outflow.  It is probably very
young, and may not even have begun a phase of mass loss yet.

The SMA spectrum of IRAS~16293$-$2422 reports the first astronomical
identification of many lines of organic and other molecules at 300 and
310~GHz.  The species detected are typical of hot cores, the emission
from which is mainly associated with source A\@.  The abundances of
second generation species, especially of sulphur-bearing molecules,
are significantly higher than predicted by chemical models for this
source to date, and we suggest that shocks are probably needed to explain
these enhancements.  The peaks in the integrated emission from molecules
having high rotation temperatures coincide with the centimeter source
A1, also highlighting the key role of shocks in explaining the nature
of hot cores.  Finally, we use the high brightness temperature of the
submillimeter dust emission from source B to demonstrate the unambiguous
detection of infall by observing redshifted SO ($7_7$$-$$6_6$) absorption
against the emission from its dust disk.

\end{abstract}


\section{Introduction}
\label{introduction}

Our understanding of the formation of binary and multiple stellar systems
in clusters lags significantly behind that of isolated star formation,
and yet this is the mode in which most stars form.  The origin of the
problem is the following: the distribution of the companion-star fraction
for T Tauri stars and field G dwarfs peaks at separations $\sim 40$--60~AU
(e.g., Patience et al.\ 2002), while the ability to resolve structure
on these size scales {\it during} the star formation process, when the
protostars are deeply embedded and invisible at optical and infrared
wavelengths, is not yet routinely available.  Distinguishing between the
possible binary formation mechanisms, such as prompt fragmentation during
collapse or the formation of instabilities in massive disks (Tohline
2002), therefore remains difficult.  Until the instruments needed to
test these theories observationally become available, we must focus on
identifying binary protostars and on understanding their interactions
with their immediate environments.

IRAS~16293$-$2422 was one of the first protostars to be identified as
a potential binary system, based on the separation of the millimeter
continuum emission into two peaks (Mundy, Wilking, \& Myers 1986),
and on the detection of centimeter radio continuum emission from each
component (Wootten 1989).  Since then it has been extensively studied
using single-dish telescopes and interferometers, and has been found
to drive two separate CO outflows (Mizuno et al.\ 1990; Stark et al.\
2004).  To date it has been assumed that each of the binary components
drives one of the outflows, although the northern binary component,
source B, exhibits very narrow linewidths and little evidence of high
velocity gas close to the source.  This latter fact led Stark et al.\
(2004) to propose that source B is a T Tauri star and that the outflow it
drove in the past is now a fossil flow, but this interpretation also has
its problems.  Thus it is not even clear yet which sources drive which
outflows.  IRAS~16293$-$2422 also exhibits strong emission from organic
and other species more typically associated with hot cores in massive
star-forming regions (e.g., van Dishoeck \& Blake 1998), especially at
the position of the southern component, source A (Bottinelli et al.\
2004; Kuan et al.\ 2004).

IRAS~16293$-$2422 was also one of the first sources for which the
detection of infall in the surrounding envelope was claimed (Walker et
al.\ 1986), based on redshifted self-absorption observed in a line of
CS\@.  The presence of the outflows and other velocity gradients across
the cloud core made the claim controversial, however (Menten et al.\
1987; Narayanan, Walker, \& Buckley 1998).  There remain, therefore,
several outstanding issues relating to the nature of the components of
the binary system in IRAS~16293$-$2422: the relative evolutionary states
of the two components; the origin of the two outflows; the nature of
the hot core emission and its relationship to the two binary components;
and the dynamics of the gas associated with the two components including
the identification of gravitational collapse.

The high spatial resolution now available with the Submillimeter Array
(SMA)\footnote{The Submillimeter Array is a joint project between
the Smithsonian Astrophysical Observatory and the Academia Sinica
Institute of Astronomy and Astrophysics, and is funded by the Smithsonian
Institution and the Academia Sinica.}, and the wide bandwidth enabling
many spectral lines to be covered simultaneously, makes this an ideal
instrument with which to try to answer some of the outstanding questions
relating to IRAS~16293$-$2422.  The association of this source with
strong, spatially-compact dust emission also raises the prospect of
excellent image quality using self-calibration techniques (Pearson \&
Readhead 1984).  The particular frequency settings chosen for the SMA
observations presented here were aimed at optimizing the possibility of
detecting the high dipole-moment molecule H$_2$CO in absorption against
the compact dust emission, as discussed in Section~\ref{infall}.

Many observations of IRAS~16293$-$2422 have also been made using the
Very Large Array (VLA) over the years, establishing one of the longest
time baselines for high-resolution centimeter radio measurements of
any source.  Here we combine new SMA data with an analysis of archival
VLA data spanning 18 years, shedding new light on the nature of the
centimeter and millimeter sources embedded in the cloud core.

IRAS~16293$-$2422 lies in the L1689 dark cloud within the $\rho$ Ophiuchi
star-forming region.  The distance to $\rho$ Oph has been reported to
be 160~pc based on optical and infrared photometry of stars associated
with the $\rho$ Oph cloud (Chini 1981), and more recently a distance of
120~pc was obtained from the reddening of optical Hipparcos and Tycho
data (Knude \& H\o g 1998).  The high extinction toward $\rho$ Oph
limits optical reddening techniques to the outer edges of the cloud.
For this reason, and to enable direct comparison with most of the
published literature on IRAS~16293$-$2422, we assume a distance of 160~pc.
However, derived quantities that depend on distance are given with $D$
explicit, and if the interpretation of any result depends sensitively
on $D$ this is discussed.

\section{Observations}

\subsection{Submillimeter wavelength SMA data}

The submillimeter observations of IRAS~16293$-$2422 were made using
the SMA located on Mauna Kea, Hawaii.  Data were obtained on 2004 June
25 and 2004 July 27, in its ``extended'' and ``compact'' configurations
respectively.  The correlator was configured to cover 2~GHz of bandwidth,
centered close to the frequency of the H$_2$CO $4_{13} \rightarrow
3_{12}$ line at 300.836635~GHz, in the lower sideband of the receiver.
The intermediate frequency of the SMA is 5~GHz, giving an upper sideband
centered 10~GHz higher.  Data are obtained from both sidebands.
The instrumental phase and amplitude variations were monitored
by observing the nearby calibrators NRAO~530 and PKS~J1625$-$2527.
Absolute flux calibration was obtained from observations of Callisto and
the quasar 3C454.3, which was assumed to have a flux density of 4.6~Jy
at 300~GHz.  The spectral response of the bandpass was obtained from
measurements of 3C454.3 and PKS~J1924$-$2914.  The uncertainty in the
absolute flux calibration is estimated to be 10\% for the continuum,
but is somewhat worse ($\sim 20$\%) for individual channels in the
spectral line data because of the signal-to-noise ratio obtained for
the bandpass calibration.

The 2~GHz total bandpass was divided into 24 overlapping correlator
segments.  Twenty-three of these were set up to have 128 channels
covering 104~MHz (channel separation 0.8125~MHz).  The other segment had
512 channels over the same bandwidth, centered on the H$_2$CO line, to
provide a spectral resolution of 0.203~MHz (0.2~km~s$^{-1}$).  The lower
sideband in particular contained several strong emission lines, so a
continuum dataset was formed from an average of the line-free channels
in the lower and upper sidebands.  This continuum emission was then
subtracted from the spectral line data.  Four of the correlator segments
at the high(low) frequency end of the lower(upper) sideband exhibited
problems, and have been excluded from the data entirely.

The data were reduced using the SMA data reduction package MIR and also
using MIRIAD and AIPS\@.  The continuum emission is strong enough for
self-calibration techniques to be used to correct for tropospheric
phase fluctuations on timescales shorter than the observations of
the phase calibrators.  The synthesized CLEAN beam using an AIPS
robust weighting of 0, intermediate between natural and uniform to
optimize both resolution and sensitivity (Briggs 1995), is $1.91''
\times 0.90''$ at P.A. 24$^\circ$, and the rms noise in the resulting
continuum image is 19~mJy~beam$^{-1}$.  This noise level is approximately
a factor of 10 times the theoretical noise expected for typical system
temperatures of 400~K, and is primarily due to the problems of imaging
sources comprising considerable extended emission using a sparse array.
The rms noise in continuum-subracted, line-free channels is approximately
3 times theoretical.  The primary beam of the SMA at the frequency of
these observations is approximately 34$''$ FWHM\@.  None of the images
presented here has had a primary beam correction applied for display
purposes, but all of the quantitative results have been obtained from
images that have been corrected for the response of the primary beam.

\subsection{Centimeter wavelength VLA data}
\label{obs_radio}

We have also retrieved all of the available continuum data for
IRAS~16293$-$2422 from the VLA archive that have sufficient resolution
to separate components A and B ($\theta_{\rm beam} \la 5''$).
Most of these have previously been reported in the literature, and
Table~\ref{archive_sum} summarizes the datasets and gives references to
the descriptions of those data; the details of the observing parameters
can be found in these references.  For all the data not previously
published, the time-dependent amplitude and phases were calibrated using
the quasars PKS~J1626$-$2951 or PKS~J1625$-$2527, and the absolute flux
density scale was obtained from measurements of 3C286.  The uncertainty in
the overall absolute flux density scale used by the VLA is estimated to be
5\% from 0.4 to 15~GHz (Baars et al.\ 1977), and 10\% at 22 and 43~GHz.
However, there is some evidence of variability in the emission from
source A from multi-epoch VLA measurements (see Section~\ref{results_var}
below), so an overall uncertainty of 10\% is also assumed for this
source at the lower frequencies where only one epoch is available,
namely, 1.5 and 5~GHz.  All the data were reduced and imaged using AIPS.

While in general our images and flux densities agree to within 2-$\sigma$
of those previously published, there is one exception.  The 15~GHz image
from Wootten (1989) shows that source A comprises two objects, denoted
A1 and A2, with A1 having a peak flux density per beam approximately
three times stronger than A2.  Our image, on the other hand, shows
that the two components have comparable peak flux densities per beam.
The difference arises because self-calibration using the 5~GHz image as a
model was applied to the data presented by Wootten.  Self-calibration can
only improve an image if there is a sufficiently high signal-to-noise
ratio in the visibility data; unfortunately, this is not the case for
the 15~GHz data, and under these circumstances false source structure
can be introduced if self-calibration is used.  Our image has not had
any self-calibration applied, and we believe it to be more reliable than
that presented by Wootten.

All the archival VLA data were observed in B1950 coordinates.  For
comparison with the SMA data all the VLA data have been precessed to
J2000.  Furthermore, the position used for one of the phase calibrators,
PKS~J1626$-$2951, has been refined and improved over the years.  All the
VLA data using PKS~J1626$-$2951 as a phase calibrator have therefore
been adjusted to match the latest position in the VLA Calibrator
Manual\footnote{Accessible from http://www.vla.nrao.edu/astro/},
of R.A. (J2000) = $16^{\rm h} 26^{\rm m} 06\fs0208$, dec.\ (J2000) =
$-29^\circ 51' 26\farcs971$.  The absolute positions for all the VLA
phase calibrators used are good to $\sim 2$~mas.

\section{Results}
\label{results}

\subsection{Continuum emission}

\begin{figure}
$$\includegraphics[scale=0.8]{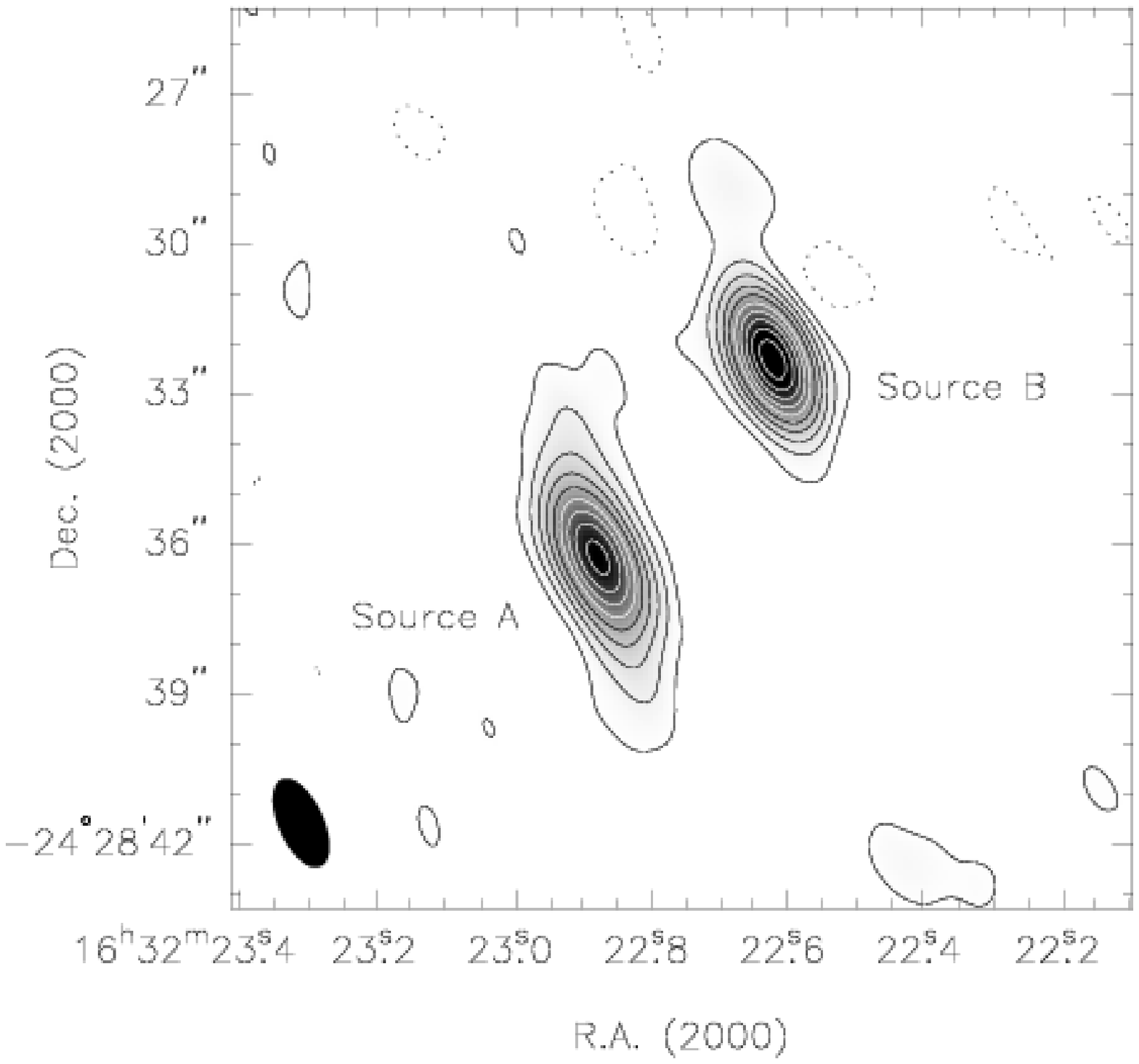}$$
\caption{Continuum image of IRAS~16293$-$2422 at $\nu = 305$~GHz
with CLEAN beam $1.91'' \times 0.90''$ at P.A. $24^\circ$, shown at
bottom left.  Contours are spaced at $-$3, 3, 8, 15, 24, 35, 48, 63,
80, 99, and 120 times the rms noise of 19~mJy~beam$^{-1}$.  The peak
1~mm flux densities for sources A and B are $2.08 \pm 0.02$ and $2.63
\pm 0.02$~Jy~beam$^{-1}$ respectively.  \label{cont_r0}}
\end{figure}

\begin{figure}
$$\includegraphics[angle=-90,scale=0.7]{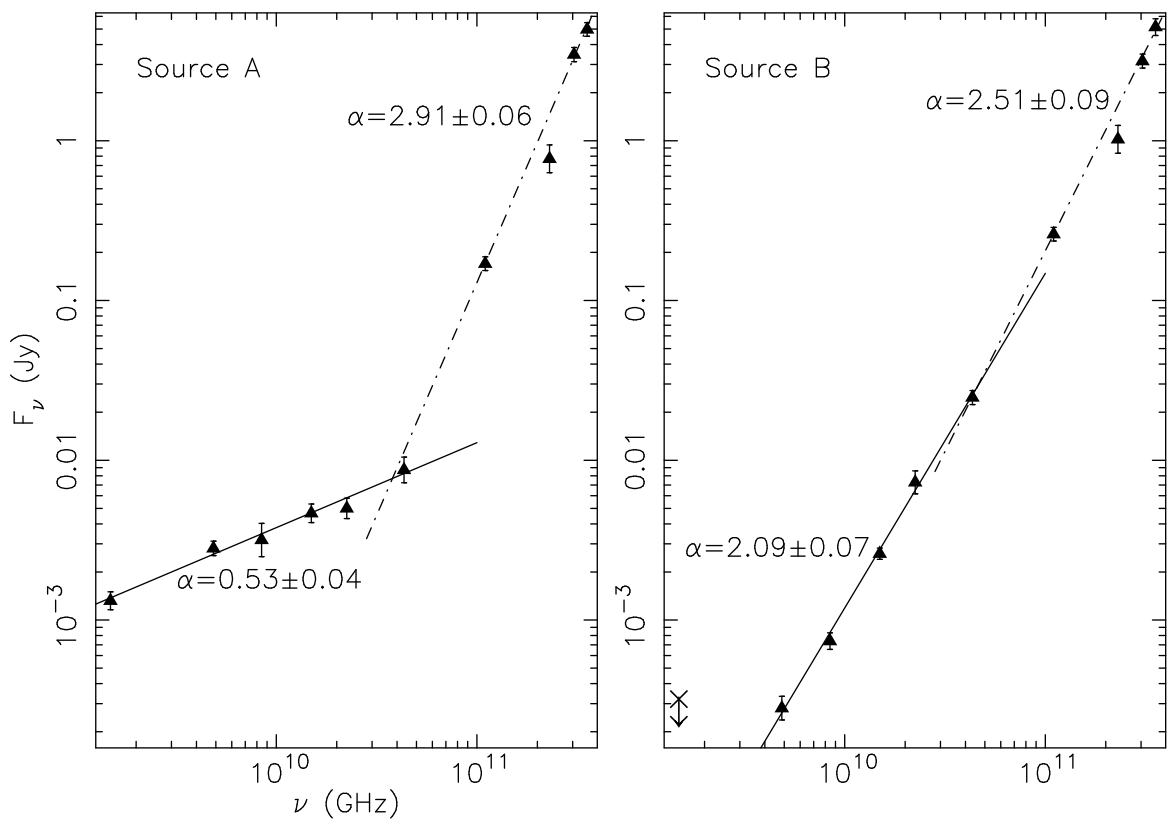}$$
\caption{Continuum spectra of sources A and B compiled from the
SMA and VLA data (Table~\ref{int_fluxes}), including data from the
literature at 110 and 230~GHz (Bottinelli et al.\ 2004) and 354~GHz
(Kuan et al.\ 2004).  The lines show best-fit power laws, $F_\nu \propto
\nu^\alpha$, to the data for $\nu < 100$~GHz (solid) and $\nu > 100$~GHz
(dot-dash).  \label{sed}}
\end{figure}

The $\lambda = 1$~mm SMA continuum image of IRAS~16293$-$2422 is shown
in Fig.~\ref{cont_r0}.  Integrated flux densities from the SMA and
archival VLA data are summarized in Table~\ref{int_fluxes}, and plotted
in Fig.~\ref{sed}.  Where multiple epochs at a particular frequency are
available, the error quoted is the standard deviation of the multi-epoch
measurements combined with the absolute uncertainty in quadrature.
Fig.~\ref{sed} also includes data points from other measurements reported
in the literature that have sufficient resolution to separate sources A
and B, at 110 and 230~GHz (Bottinelli et al.\ 2004) and at 354~GHz (Kuan
et al.\ 2004).  Power-law fits to the interferometric data, $F_\nu \propto
\nu^\alpha$, are also plotted for $\nu < 100$~GHz and $\nu > 100$~GHz.
The fact that sources A and B are embedded in extended emission from
the surrounding envelope in fact makes the spectral indices derived
from formal fits to the $\nu > 100$~GHz data rather more uncertain
than the errors given in Fig.~\ref{sed}, since no attempt has been
made here to match the $u$-$v$ coverage at each frequency.  Indeed,
that this is a potential problem can clearly be seen for the 230~GHz
flux density from Bottinelli et al.\ (2004), for which the shortest
baseline is a factor of $\sim 2$ longer than that of the 110~GHz data.
This is also illustrated in the 110~GHz images reported by Looney,
Mundy, \& Welch (2000).  The 230~GHz data have therefore not been
included in the fit.  Interpolating between the 1.3~mm and 850~$\mu$m
total integrated single-dish flux densities from Sch\"oier et al.\ (2002)
using a power law we expect a total 1~mm flux density of 18~Jy; the SMA
is therefore filtering out approximately 60\% of the continuum emission
associated with the extended envelope.  At 43~GHz the largest angular
scale to which the VLA is sensitive in the A-configuration is 1.3$''$.
The 43~GHz measurements available to date seem to be sensitive only to
the compact emission components also responsible for the emission at
lower frequencies, and are missing flux associated with the more extended
emission detected at higher frequencies.

\begin{figure}
\plottwo{fig3a.ps}{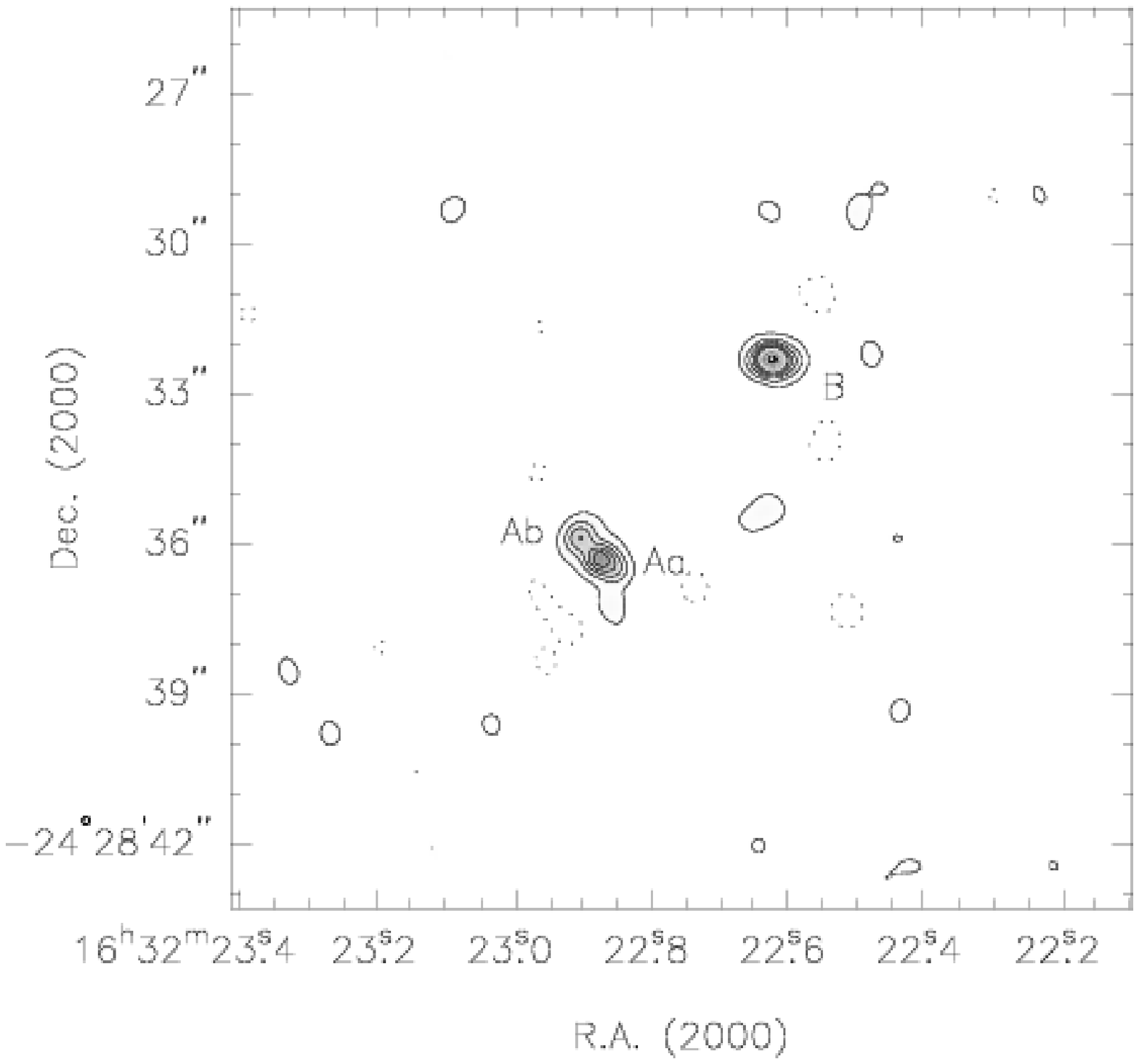}
\caption{{\it Left:} plot of visibility amplitude against $u$-$v$ distance
projected along a position angle of $-45^\circ$.  {\it Right:} Continuum
image obtained using only data $> 55$~k$\lambda$, super-uniform weighting,
and a circular restoring CLEAN beam of $0.4''$ FWHM\@.  Contours are at
the same levels as in Fig.~\ref{cont_r0}.  \label{cont_su}}
\end{figure}

The 1~mm continuum emission detected by the SMA from sources A and B
comprises both extended components and more compact emission.  In order
to assess the relative contributions from both, we have made an image
using only the long baseline data (a lower limit of 55~k$\lambda$
was chosen from inspecting plots of visibility flux versus $u$-$v$
distance, the point at which the circumbinary envelope starts to
contribute significantly to the overall flux: see Fig.~\ref{cont_su}),
increasing the relative weights of the longest baselines (super-uniform
weighting), and restoring the image with a CLEAN beam of $0.4''$,
slightly smaller than half the fringe spacing on the longest baseline,
$0.54''$ (Fig.~\ref{cont_su}).  This image essentially illustrates
the location of the CLEAN components derived from the deconvolution
of the point spread function, as obtained from the longest baselines
present in the data.  It is apparent in this ``super-resolution'' image
that while the compact emission from source B remains singly-peaked,
source A comprises a compact source close to the position of the overall
continuum peak in Fig.~\ref{cont_r0}, along with a second, new source
offset by 0.64$''$ to the northeast.  This object does not coincide
with any of the previously-identified sources in this region (the
closest is the centimeter radio source A1, which is $\sim 0.4''$ to the
southwest), and contains approximately 30\% of the flux associated with
the compact emission from source A in Fig.~\ref{cont_su}.  We interpret
this new source as a candidate protostar.  In what follows we denote the
stronger of the two components Aa, and the new candidate companion Ab.
Table~\ref{compact} gives positions and flux densities for the compact
emission components shown in Fig.~\ref{cont_su}.  The fraction of the
flux detected by the interferometer associated with the compact emission
shown in Fig.~\ref{cont_su} for source A is $\sim 60$\%, and for source
B is $\sim 90$\%.

Since source A is now shown to comprise multiple submillimeter
and centimeter radio components we use ``source A'' to refer to the
combination of all components, Aa+Ab+A1+A2.  Discussion of an individual
component will refer directly to the name of that component.

\subsubsection{Proper motion and the relationship between the centimeter
and submillimeter sources}

\begin{figure}
$$\includegraphics[angle=-90,scale=0.73]{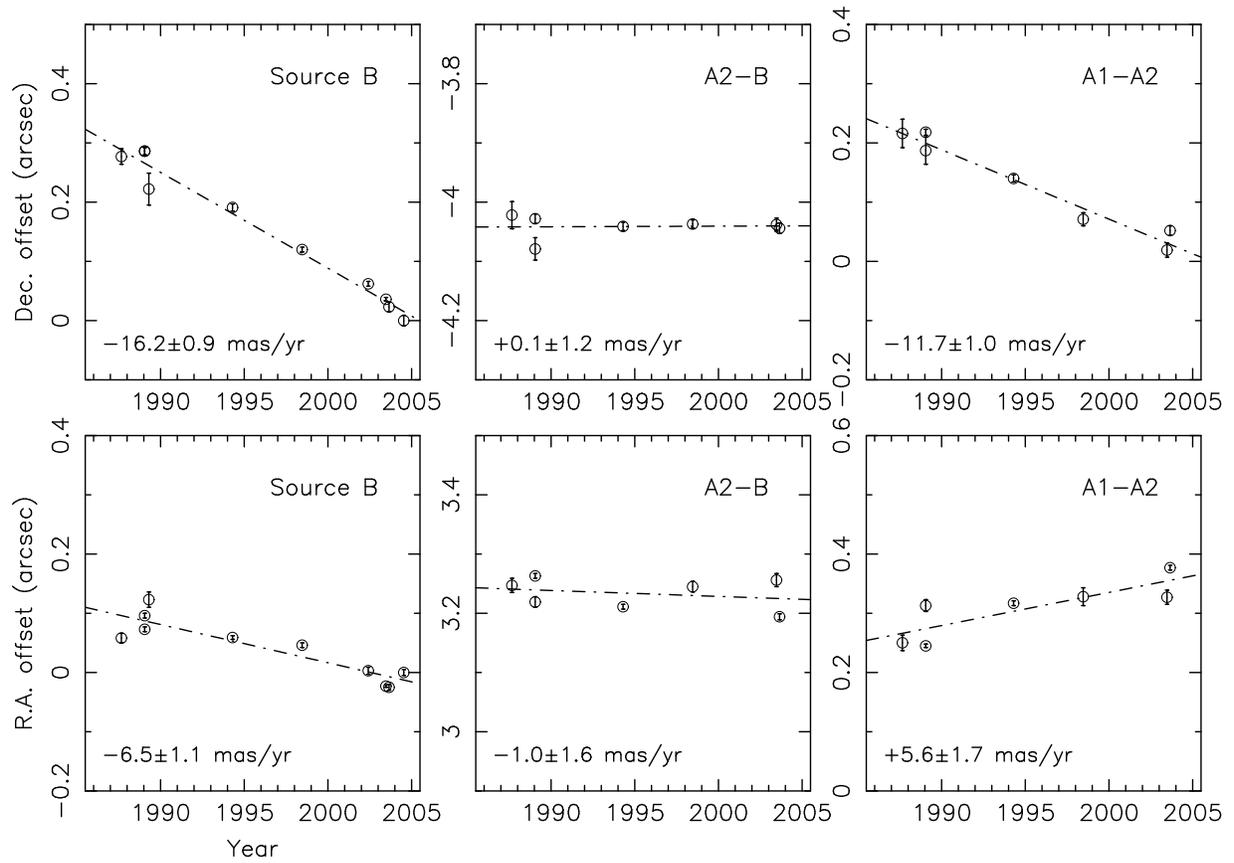}$$
\caption{{\it Left:} the proper motion of source B, plotted as
offsets from the submillimeter position given in Table~\ref{compact}.
{\it Center:} the motion of radio source A2 relative to source B, as
a function of time.  {\it Right:} the separation A1$-$A2 as a function
of time.\label{pmotions}}
\end{figure}

In order to investigate the relationship between the submillimeter
emission and the radio continuum sources, we first need to account for
the proper motion of the sources.  The proper motion of sources A and B
has been reported previously by Loinard (2002), based on two epochs of
VLA 8~GHz data (1989 and 1994), and indicate motion presumably associated
with the L1689N cloud in which sources A and B are embedded (see also
Curiel et al.\ 2003).  When observed at high resolution the centimeter
radio emission from source A comprises two components, denoted A1 (to
the east) and A2 (to the west) by Wootten (1989), separated by 0.35$''$.
We show below that the overall structure of the centimeter radio emission
from source A has been changing with time, so the simpler structure of
source B makes measurement of its proper motion more straightforward.
In Fig.~\ref{pmotions} we plot the position of source B derived from the
SMA data, and all the VLA data having a resolution of 1$''$ or better,
as a function of time, along with the offset between sources A2 and B,
and between A1 and A2, for those measurements which resolve the two
components of source A\@.  Fig.~\ref{pmotions} demonstrates that the
proper motion of sources A2 and B are in common, and that their motion is
primarily in declination.  The overall proper motion of sources A2 and
B, 17~mas~yr$^{-1}$, corresponds to $13 (D/{\rm 160~pc})$~km~s$^{-1}$.
Fig.~\ref{pmotions} also illustrates significant motion of A1 relative
to A2, amounting to $10 (D/{\rm 160~pc})$~km~s$^{-1}$.

\begin{figure}[ht]
$$\includegraphics[angle=-90,scale=0.6]{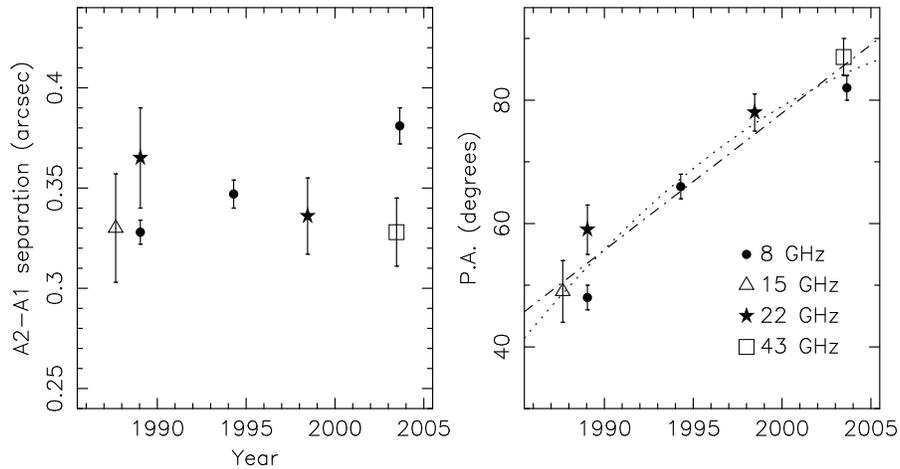}$$
\caption{Projected separation $|$A1$-$A2$|$ {\it (left)} and the position
angle of the vector joining A2 and A1 {\it (right)} as a function of time.
The lines represent the best linear (dot--dash) and quadratic (dotted)
fits to the position angle data.  The linear fit gives $2.22 \pm
0.26$~deg~yr$^{-1}$.  \label{pangle}}
\end{figure}

We have corrected for the proper motions described above, and other
uncertainties in the absolute positions caused by using different
quasars for the phase calibration for some of the measurements,
by shifting all the archival VLA to align source B with the 1~mm
position in Table~\ref{compact}.  Fig.~\ref{overlays} presents the
result of overlaying the shifted radio continuum emission on the
1~mm super-resolution continuum image.  When observed with sufficient
resolution to separate the two centimeter radio sources A1 and A2 it is
clear first that their orientation has been changing over the years as
indicated in Fig.~\ref{pmotions}, and second that they lie on either side
of the submillimeter continuum source Aa.  In Fig.~\ref{pangle} we plot
the projected separation between A1 and A2, and their position angle,
as a function of time.  Considering only the 8~GHz data we might conclude
that the separation between A1 and A2 has been increasing slightly, and
that the rate at which the position angle is changing is slowing down.
However, this conclusion becomes marginal when the results at other
frequencies are included.  Fig.~\ref{pmotions} shows that a linear
fit to the position offsets in R.A. and dec.\ matches the data well.
The projected separation between A1 and A2 has remained constant, at
$0.35'' \pm 0.02''$, while the position angle is equally well-fitted by
a straight line as by a second order polynomial function.  The quadratic
function shown in Fig.~\ref{pangle} (dotted line) turns over in 2014, so
regular monitoring will be needed over the next few years to establish
whether the rate of change in position angle is indeed slowing down,
as suggested by the three 8~GHz points.

Since the 1~mm SMA data are not able to resolve the separation of the
two centimeter radio sources, it is important to establish whether the
apparent location of Aa between A1 and A2 in the super-resolution image is
real, or whether the location of the CLEAN components just corresponds to
the centroid of submillimeter emission associated directly with A1 and A2.
To test this we have made simulated images comprising a source at Ab, and
sources at the positions of A1 and A2 from epoch 2003.65 (the most recent
VLA data) with a total flux that of Aa given in Table~\ref{compact}.
We find that for a flux ratio $F_{\rm A2}/F_{\rm A1}$ of unity the peak
is indeed located at the observed position of Aa in Fig.~\ref{cont_su}.
For a flux ratio of 1.8, the same as that observed at 8~GHz for this
epoch, the peak position would be offset by 0.12$''$, which is more
than 6-$\sigma$ away from that measured.  We therefore conclude that
if the centimeter radio sources A1 and A2 are each associated with
dust continuum emission, their submillimeter flux ratio is close to unity.
The resulting implications for the nature of Aa, A1, and A2 is discussed
further in Section~\ref{discussion}.

\subsubsection{Source variability}
\label{results_var}

\begin{figure}
$$\includegraphics[scale=0.63,angle=-90]{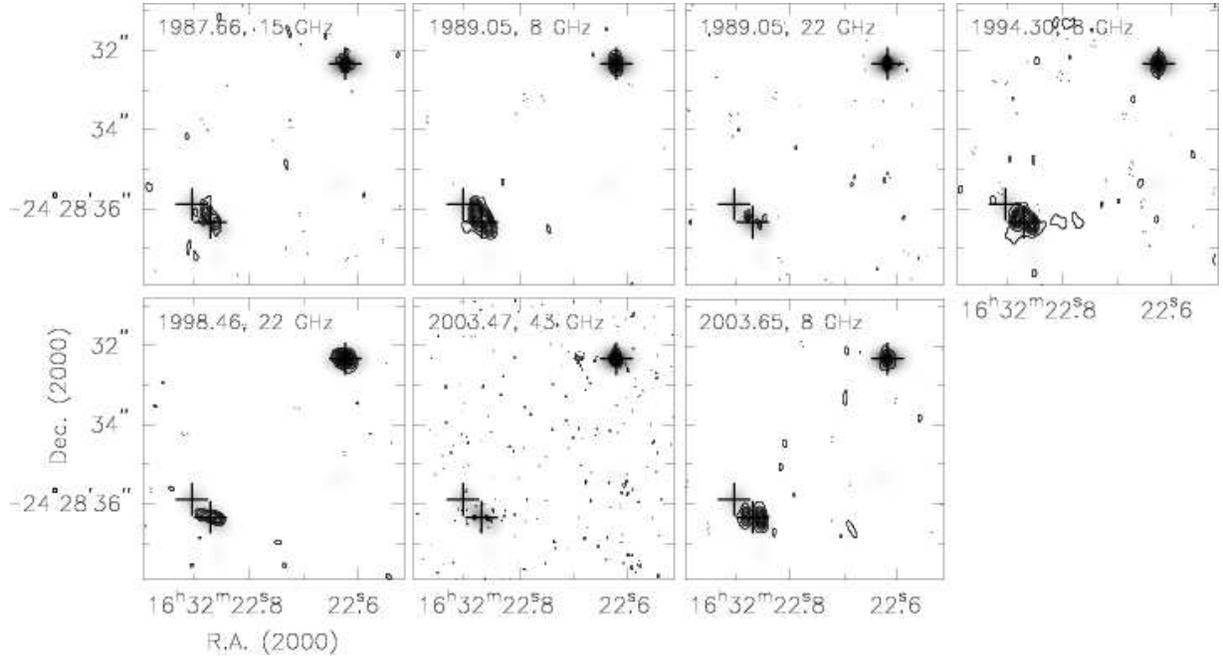}$$
\caption{VLA data (contours) overlaid on the 1~mm super-resolution image,
after aligning source B in the different epochs.  The crosses indicate
the positions given in Table~\ref{compact}.  The size of the crosses
are not proportional to the positional accuracy of the submillimeter
sources.  \label{overlays}}
\end{figure}

\begin{figure}
$$\includegraphics[angle=-90,scale=0.7]{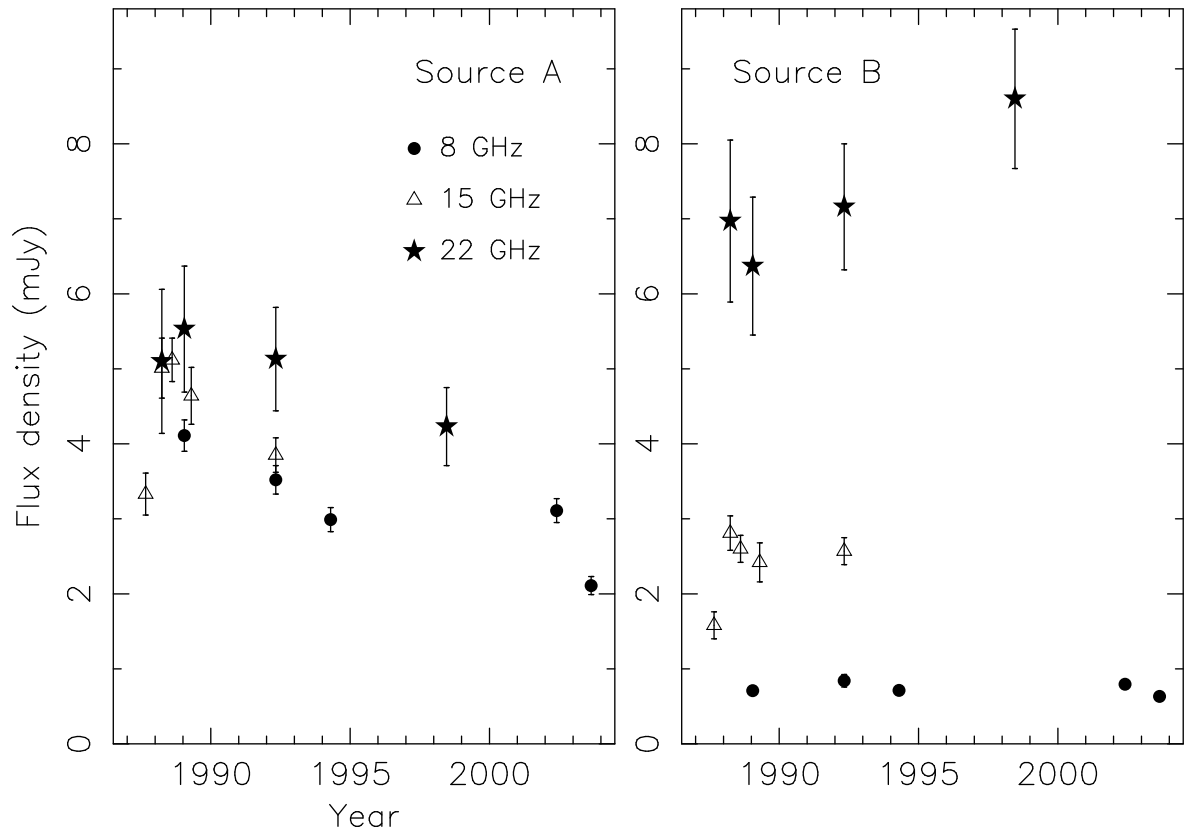}$$
\caption{Integrated fluxes for sources A {\it (left)} and B {\it (right)}
as a function of time.  \label{fluxvar}}
\end{figure}

Along with the clear secular structural evolution of the radio emission
from source A shown in Fig.~\ref{overlays}, there is also evidence
for flux variability.  Fig.~\ref{fluxvar} illustrates the integrated
flux density for sources A and B as a function of time at 8, 15, and
22~GHz, where the error bars include the absolute flux uncertainty
described in Section~\ref{obs_radio}.  At all three frequencies, source
B is consistent with being nonvariable, apart from the earliest 15~GHz
measurement which comes from the data first presented by Wootten (1989).
These data suffered significantly from poor tropospheric phase stability,
and an analysis of images of the phase calibrator for this epoch suggests
that the observed flux densities for IRAS~16293$-$2422 are $\sim 20$\% too
low due to the decorrelation.  We have not, therefore, included the flux
density measurements from this epoch in the average 15~GHz flux densities
plotted in Fig.~\ref{sed} for sources A and B, and exclude it from our
discussion of variability here.  Thus we conclude that the flux density
of source B is constant in time, at all three frequencies measured.

Source A, on the other hand, clearly exhibits a trend of decreasing
flux density between 1988 and 2003 at both 8 and 15~GHz, while the
spectral index remains approximately constant.  The observations able
to separate the centimeter radio components show that the emission from
both A1 and A2 has been decreasing, but that A1 is fading faster than
A2, with $F_{\rm A2}/F_{\rm A1} \sim 0.7$ around epoch 1990, compared
to $\sim 1.8$ in 2003.  During this time the position angle of the
vector joining A1 and A2 has traversed the position angle of the vector
joining the submillimeter components Aa and Ab (see Fig.~\ref{overlays}).
It is possible that the dramatic variability of A1 has been produced
by the interaction of a wind or jet with Ab and/or dense gas in its
immediate environment, or from intrinsic variability of the source of
ionized gas.  The origin and properties of this plasma is discussed
further in Section~\ref{continuum_origin}.

Possible short term variability of the 15~GHz radio emission has
been investigated by Estallela et al.\ (1991) between 1987 and 1988.
No evidence for variability on timescales of several hours to a year was
found by those authors, but this is consistent with the variability of
source A on longer timescales, as demonstrated by Fig.~\ref{fluxvar}.

\subsection{Spectroscopy}

All the molecular line emission was imaged with an AIPS robust
weighting parameter of 2, close to natural weighting, to give the
best sensitivity.  The resulting synthesized CLEAN beam is $2.45''
\times 1.22''$ at P.A. $31^\circ$ for both lower and upper sidebands.
Fig.~\ref{long_spec} shows the full lower and upper sideband spectra
for sources A and B, averaged over boxes approximately $1.5'' \times
1.5''$ in size.  Only one other spectrum covering $\nu \sim 300$~GHz has
been reported in the literature, a low spectral resolution (200~MHz)
observation of the Orion molecular cloud core by Serabyn \& Weisstein
(1995).  Fig.~\ref{long_spec} therefore shows the first astronomical
detection of many of the lines indicated.  Line identifications have been
obtained from a comparison with the spectral line catalogs of Pickett
et al.\ (1998) and M\"uller et al.\ (2001).  Line identifications are
particularly difficult for the weak transitions of organic species, for
which many potential blends are possible within a typical linewidth of a
few MHz.  Firm identifications of organic molecules have therefore only
been made when all transitions lying within the lower or upper sideband
are observed with the expected relative line strengths.  The main organic
species that have been detected are dimethyl ether (CH$_3$OCH$_3$), A-
and E-type methyl formate (CH$_3$OCHO), cyanoacetylene (HC$_3$N), methanol
(CH$_3$OH), and formaldehyde (H$_2$CO).  A possible line of Si$^{34}$S
may be blended with methyl formate emission.  Other significant species
detected are the sulphur-bearing molecules NS, SO$_2$, H$_2$S, and SO,
and the deuterated species DNO (in absorption) and HDO\@.  Several
lines remain unidentified, emphasizing the need for more complete line
lists for complicated organic molecules at submillimeter wavelengths.
Table~\ref{line_ids} summarizes the transitions of all molecules observed,
and gives integrated line fluxes for positively-identified lines not
blended with emission from other species, and also for H$_2$CO, which
is blended with methyl formate emission.

\begin{figure}
$$\includegraphics[scale=0.7,angle=-90]{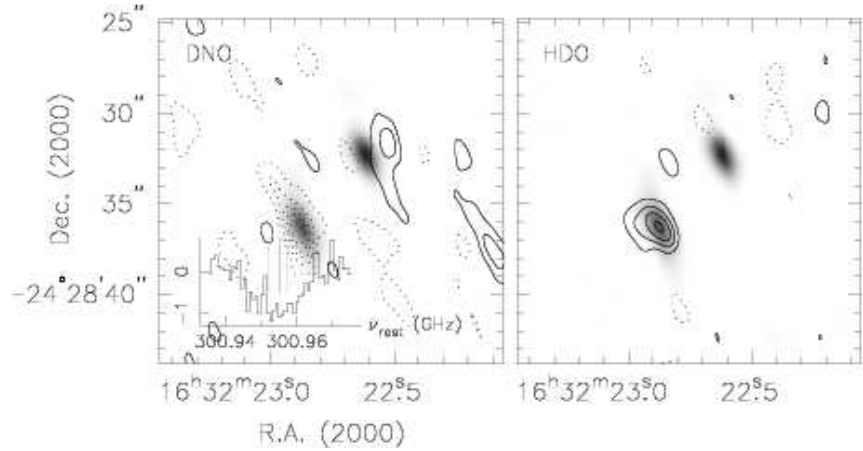}$$
\caption{Emission and absorption from deuterated species.  Contours are
linearly spaced at approximately 2$\sigma$ intervals of 2.80 (DNO),
and 2.64 (HDO) Jy~km~s$^{-1}$.  Dotted contours are negative.  DNO is
detected in absorption toward source A, and its unsmoothed spectrum,
with the location of DNO lines indicated, is in the bottom left of
the DNO panel.  \label{deuterated}}
\end{figure}

\begin{figure}
$$\includegraphics[scale=0.9,angle=-90]{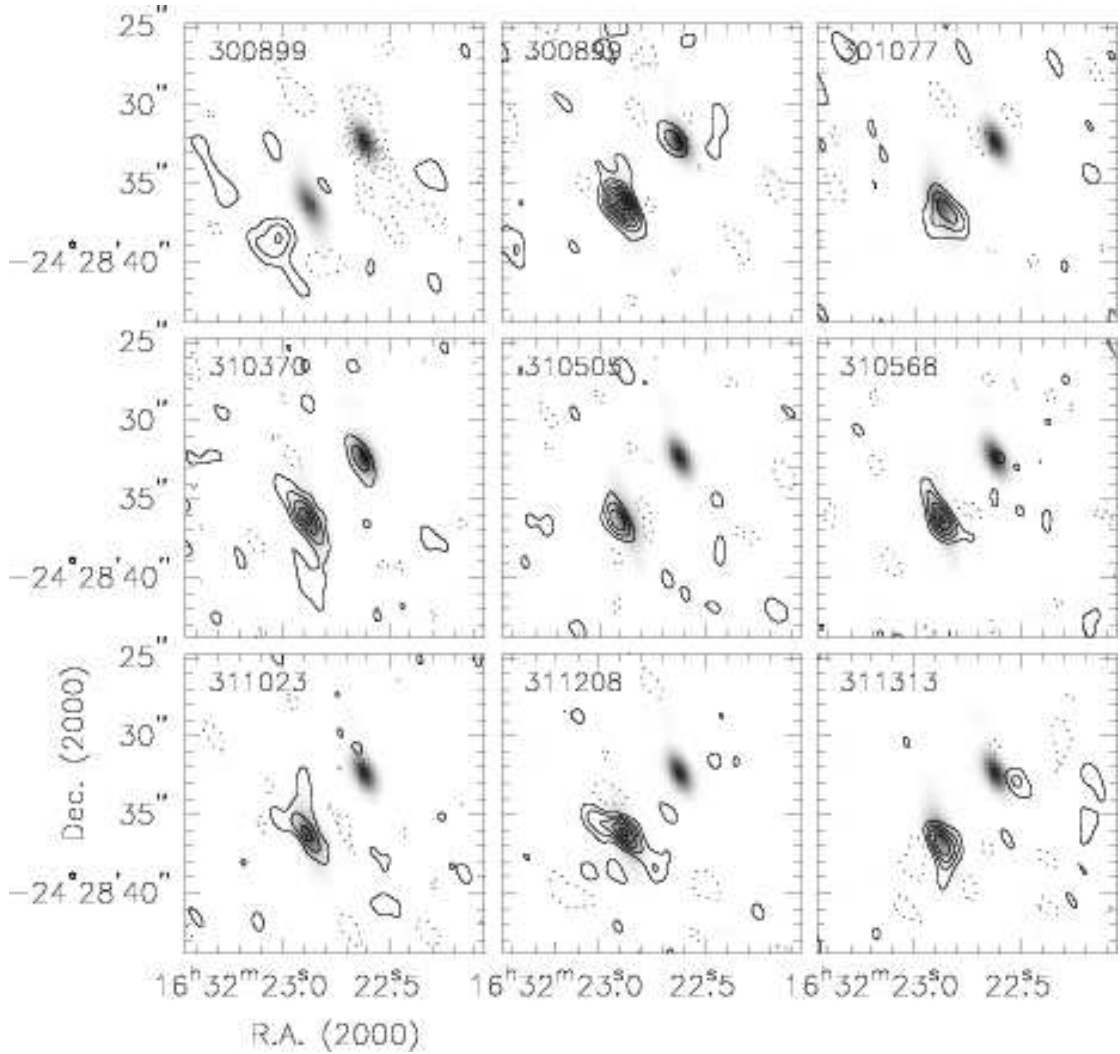}$$
\caption{Emission from unidentified lines.  Each panel is marked with
the approximate center rest frequency of the line in MHz.  In the case
of the 300899~MHz line, there is a feature detected in absorption toward
source B to the redshifted side of the strong emission line observed
toward source A\@.  It is possible that these originate from completely
different lines, but in the absence of further information they have
both been designated as 300899~MHz here.  All contours are linearly
spaced at approximately 2$\sigma$ intervals.  \label{unidentified}}
\end{figure}

Lines detected at 6$\sigma$ or better in integrated emission
(corresponding approximately to a flux density cutoff of
0.7~Jy~beam$^{-1}$ in Fig.~\ref{long_spec}) are displayed in
Figs.~\ref{organic}--\ref{unidentified}, showing that the molecular line
emission is centered close to source A, with some species also exhibiting
emission coincident with source B\@.  Unlike the results for H$_2$CCO and
$c$-C$_3$H$_2$ reported by Kuan et al.\ (2004), none of the molecules we
detect in emission are associated only with source B and not source A\@.
There is, however, a significant detection of absorption toward source
B in an unidentified line.

Single-dish observations have indicated the presence of three separate
physical and chemical components in IRAS~16293$-$2422: a cold, outer
envelope with $T \sim 10$--20~K, a circumbinary envelope with $T \sim
40$~K approximately $10''$ in size, and a compact, warm component ($T \ga
80$~K) only a few arcsec in extent (van Dishoeck et al.\ 1995; Ceccarelli
et al.\ 2000).  Of course, there is probably a continuous temperature
gradient throughout the envelope, with these size scales representing
temperatures at which molecules are liberated from the surface of dust
grains, and where ice mantles evaporate (Sch\"oier et al.\ 2002; Doty,
Sch\"oier, \& van Dishoeck 2004).  Subsequent gas-phase chemistry can
then cause significant jumps in the abundances of certain species.
Further, gas phase abundances, particularly of sulphur-bearing species,
can be dramatically changed by the passage of a shock (Wakelam et al.\
2004a; Viti et al.\ 2001).

The presence of these different regimes in the envelope mean that
single-dish measurements of beam-averaged column densities, and
their conversion to abundances through comparison with (single-dish)
beam-averaged H$_2$ column densities, will not give good estimates of the
{\it local} abundance.  To minimize this problem for our interferometer
data we do the following.  Molecular column densities have been calculated
by integrating over emission regions having a signal-to-noise ratio $\ge 2
\sigma$, using rotation temperatures derived from rotation diagrams (cf.\
Goldsmith \& Langer 1999) either reported previously in the literature,
or from new analyses presented below.  The corresponding molecular
hydrogen column densities for determining local abundances are obtained
from a 1~mm SMA dust continuum image made with the same weighting and beam
size as the molecular line data, and masked to match the extent of the
molecular emission.  We adopt a 300~GHz dust opacity $\kappa_{\rm 300~GHz}
= 1.5$~cm$^2$~g$^{-1}$, a gas to dust ratio $M_{\rm g}/M_{\rm d} = 100$,
and $T_{\rm dust} = 40$~K, as described in Section~\ref{continuum_origin}.

The H$_2$ column density derived from the dust emission is inversely
proportional to the assumed temperature, but since the line of sight
towards sources A and B probably includes emission from dust at $\sim
20$ to 80~K, a value of 40~K will not introduce errors of more than a
factor of 2 or 3.  Other sources of uncertainty are likely to dominate
the derived abundances, such as the uncertainty in the absolute value of
the dust opacity, the uncertainty in the assumed rotation temperatures,
the presence of temperature gradients in the emission regions, and the
assumption of local thermodynamic equilibrium.  Table~\ref{abundances}
summarizes the derived or assumed rotation temperatures, and the average
column densities and abundances for each molecule.  The abundances
for sources A and B cannot be separated due to the dependence of the
rotation temperature estimates on single dish data: even the new estimates
presented below use supplemental single dish data from the literature.
Thus, only a composite abundance is listed in Table~\ref{abundances}.
Instances where there is circumstantial evidence of a significant
difference between the two sources are noted below in comments on the
individual molecular species.

In the notes below we also compare the derived abundances with previous
attempts to model or predict chemical abundances in IRAS~16293$-$2422 by
Sch\"oier et al.\ (2002) and Doty et al.\ (2004); the model abundances
from these studies are also presented in Table~\ref{abundances}.
The Sch\"oier et al.\ (2002) model is based on a detailed analysis of the
radiative transfer of the observed single dish continuum and molecular
line emission from IRAS~16293$-$2422.  This model uses the derived density
and temperature profiles as a function of radius from analysis of the
continuum data to determine empirically the best fit to the molecular line
data by varying the abundances as a function of radius.  In contrast Doty
et al.\ (2004) use the density profile derived by Sch\"oier et al.\ (2002)
and the UMIST chemical network database (Millar, Farquhar, \& Willacy
1997) to determine directly the temperature profile and the abundances as
a function of time and radius.  In order to compare these models with our
SMA data we only consider the abundances reported by Sch\"oier et al.\
and Doty et al.\ on small size scales, corresponding to $r<2.5\times
10^{15}$~cm (i.e., $1\arcsec$ at 160 pc).  Further comparison of these
model results and our data are presented in Section~\ref{chem_compare}.

\subsubsection{Organic species}

\begin{figure}
$$\includegraphics[scale=0.7,angle=-90]{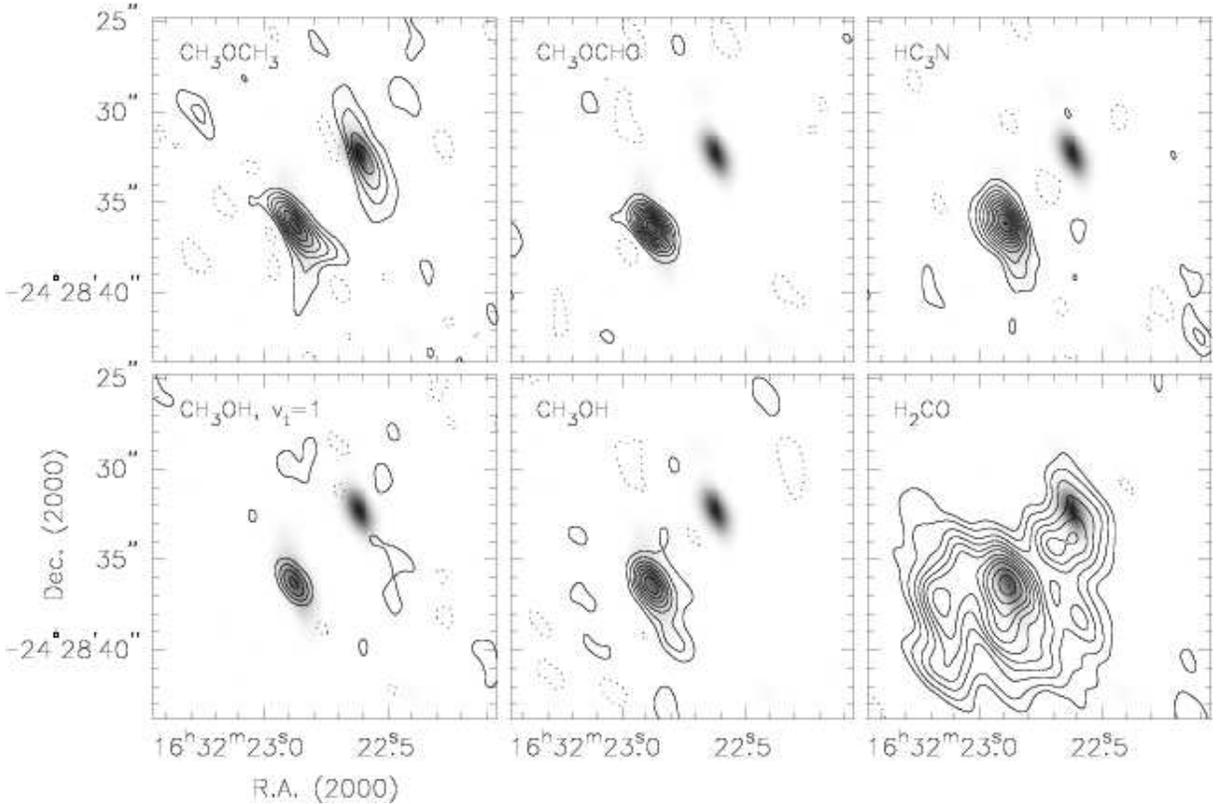}$$
\caption{Location of emission from organic species detected towards
IRAS~16293$-$2422.  The emission from CH$_3$OCH$_3$ is the sum of both
the 5$-$4 and 8$-$7 transitions in Table~\ref{line_ids}, and the
CH$_3$OCHO image is made from the sum of all A- and E-type lines not
blended with other species.  For all images apart from H$_2$CO the
contours are linearly spaced at approximately 2$\sigma$ intervals
of 2.43 (CH$_3$OCH$_3$), 1.97 (CH$_3$OCHO), 1.96 (HC$_3$N), 1.53
(CH$_3$OH, $v_t=1$), and 2.34 (CH$_3$OH) Jy~km~s$^{-1}$.  For H$_2$CO
the contours are at $-$1, 1, 2, 3.5, 5, 6.5, 8, 10, 12, 14, 17, and 20
times 2.04~Jy~km~s$^{-1}$.  \label{organic}}
\end{figure}

The compact, warm component of the envelope shows many features in common
with the hot cores usually associated with more massive star-forming
regions, albeit on a much smaller size scale.  In such hot cores,
the hydrogenated ``first-generation'' molecular species released from
ices on the dust grains, such as H$_2$CO and CH$_3$OH, then form the
basis of further gas-phase chemical reactions to form more complex
``second-generation'' species, such as CH$_3$OCH$_3$ and CH$_3$OCHO (e.g.,
Charnley, Tielens, \& Millar 1992).  Initial attempts to detect these
second-generation, complex organic molecules toward IRAS~16293$-$2422
using single-dish telescopes provided only upper limits (van Dishoeck
et al.\ 1995), and Sch\"oier et al.\ (2002) interpreted this to mean
that the time needed to form these molecules is not available on such
small scales in an infalling envelope close to a low-mass protostar.
However, very compact emission from several complex organic species
has now been reported by Bottinelli et al.\ (2004) and Kuan et al.\
(2004), and is also shown in Fig.~\ref{organic}, suggesting that the
earlier non-detections were due to beam dilution.  A range of emission
structure is exhibited by the organic molecules detected in our SMA study
(see Fig.~\ref{organic}), from the unresolved emission from a high-energy
transition of torsionally-excited methanol, to extended emission from
formaldehyde.

\subsubsubsection{CH$_{\it 3}$OCH$_{\it 3}$}

\begin{figure}
$$\includegraphics[scale=0.65,angle=-90]{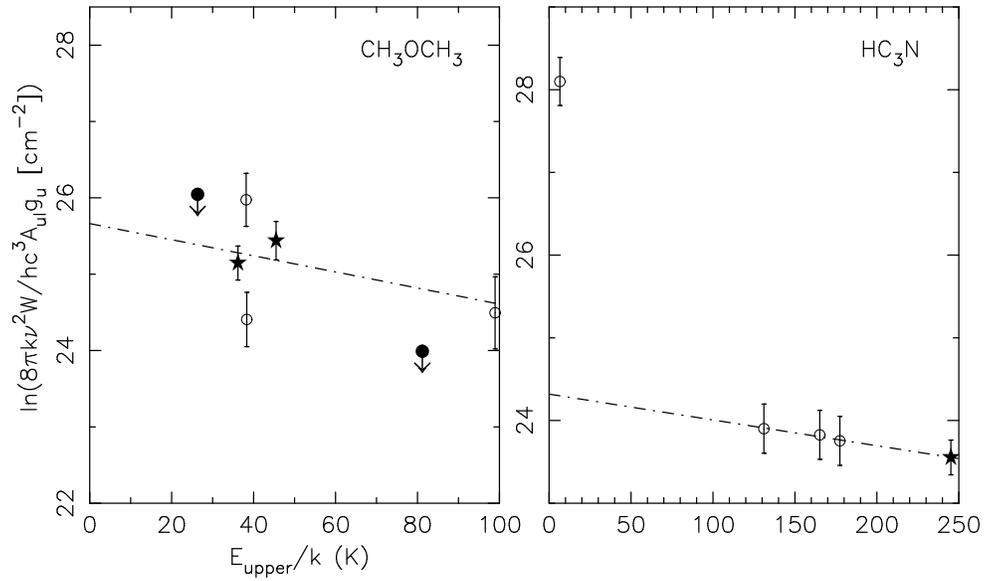}$$
\caption{Rotation diagrams for the organic species detected; data from
this work are marked as filled stars.  Supplementary CH$_3$OCH$_3$
data come from Cazaux et al.\ (2003), and the upper limits are from
van Dishoeck et al.\ (1995).  The dot-dash line is for $T_{\rm rot}
= 95$~K\@.  Supplementary HC$_3$N data are from Suzuki et al.\ (1992),
van Dishoeck et al.\ (1995), and Sch\"oier et al.\ (2002).  The dot-dash
line is for $T_{\rm rot} = 320$~K, and does not include the low energy
$J$=5$-$4 line in the fit.  \label{rot_organic}}
\end{figure}

Dimethyl ether is associated with both sources A and B\@.  Cazaux et
al.\ (2003) have also detected emission from 7$-$7, 8$-$7, and 14$-$13
transitions using the IRAM 30m telescope, and we plot a rotation diagram
including these data in Fig.~\ref{rot_organic}, from which the rotation
temperature and total column density listed in Table~\ref{abundances}
is derived.  Chemical models require high gas-phase abundances of
methanol to produce dimethyl ether (Millar, Herbst, \& Charnley 1991),
and timescales $\sim 10^4$--$10^5$~yr.  The abundance measured here,
$\sim 8 \times 10^{-8}$, is similar to that predicted by models for
the Orion compact ridge, where the grain mantles are rich in methanol
(Charnley et al.\ 1992; Caselli, Hasegawa, \& Herbst 1993).  However,
it is significantly in excess of that predicted by the physical-chemical
modeling of IRAS~16293$-$2422 by Doty et al.\ (2004), who obtain a
maximum value of $\sim 3 \times 10^{-9}$.  The Sch\"oier et al.\ (2002)
model predicts a dimethyl ether abundance of $< 4\times 10^{-8}$, within
a factor of two of the SMA value.

\subsubsubsection{CH$_{\it 3}$OCHO}

The large number of transitions detected from A- and E-type methyl formate
are unfortunately severely blended throughout the spectrum, and we have
made no attempt to separate them to provide independent column density and
abundance measurements.  The image shown in Fig.~\ref{organic} comprises
all the emission from both A- and E-type methyl formate not blended with
other species.  Note that unlike the methyl formate images presented by
Bottinelli et al.\ (2004) and Kuan et al.\ (2004), we do not detect any
methyl formate emission associated with source B\@.  In the case of the
Bottinelli et al.\ result this may be explained by the fact that these
authors observed somewhat lower energy transitions than those presented
here, consistent with the gas excitation temperatures (and CH$_{\it
3}$OCHO column density) at source B being lower than for source A\@.
However, the methyl formate transitions detected by Kuan et al.\ toward
source B are {\it higher} in energy, which is difficult to reconcile
with the upper limit of our non-detection (about a factor of two less
than their detection).  The formation of methyl formate is thought to
arise from chemical reactions involving either formaldehyde or methanol
when these species are evaporated from the surface of dust grains (e.g.,
Horn et al.\ 2004) both of which are primarily associated with source A
(Fig.~\ref{organic}).

\subsubsubsection{HC$_{\it 3}$N}

The emission from the relatively high energy level transition of HC$_3$N
is compact and associated with source A, and the rotation diagram for
this molecule (Fig.~\ref{rot_organic}) also indicates high temperatures
on these size scales (Table~\ref{abundances}).  However, there is
clearly also emission from the more extended, cooler, envelope, and
the $J$=5$-$4 flux reported by Suzuki et al.\ (1992) lies considerably
above the best-fit $T_{\rm rot} \sim 320$~K obtained using only the
higher-lying transitions.  The abundance derived, $1 \times 10^{-9}$,
is in good agreement with those obtained for the inner envelope from
an analysis of single-dish data (Sch\"oier et al.\ 2002), but is an
order of magnitude higher than that predicted by Doty et al.\ (2004)
for small size scales, of $\sim 1 \times 10^{-10}$.

\subsubsubsection{CH$_{\it 3}$OH}

Methanol is detected in both its ground and first torsionally-excited
states.  The ground state emission is fairly compact with a slight NE-SW
extension.  The emission from the $v_t=1$ line is unresolved, but offset
by +0.1$''$ in R.A. from the continuum position of Aa, close to A1.
This line has the highest energy of any of the transitions observed,
with $E_{\rm upper} = 732.4$~K\@.  Since only one transition of the
torsionally-excited state was observed we regard this detection as
tentative until other lines have been identified.

The rotation temperature derived from many lines of CH$_3$OH by van
Dishoeck et al.\ (1995) has been used in calculating the column density
and abundance from the ground and torsionally-excited state emission
in Table~\ref{abundances}.  Both the column densities and abundances
are in good agreement with the models of Sch\"oier et al.\ (2002) and
Doty et al.\ (2004).

\subsubsubsection{H$_{\it 2}$CO}

The formaldehyde emission is very extended and originates from both
the warm, compact, inner envelope, and the $\sim 40$~K circumbinary
envelope.  The peak line brightness temperature is 42~K, indicating that
especially toward source A the emission has significant optical depth.
The H$_2$CO emission is unfortunately blended with several lines of methyl
formate that fall on the low frequency side of the formaldehyde line.
Using the column densities derived by Cazaux et al.\ (2003) for methyl
formate (and noting that these authors assumed a size of 2$''$ for
the emitting region), rotation temperatures between 65 and 100~K give
potential contributions to the integrated flux in the formaldehyde line
reported in Table~\ref{line_ids} of 5--10\%.  We have subtracted 5\%
(corresponding to $T_{\rm rot} \sim 65$~K) from the integrated flux,
and use the rotation temperature derived by van Dishoeck et al.\
(1995) of 80~K for deriving H$_2$CO column densities and abundances in
Table~\ref{abundances}.  An abundance of $\sim 1 \times 10^{-7}$ matches
those predicted on small size scales by the Sch\"oier et al.\ and Doty
et al.\ models to within a factor of 2.  The column density also matches
the models reasonably well, when it is taken into account that there is
considerable extended formaldehyde emission from the envelope that has
been resolved out by the interferometer (see also Sch\"oier et al.\ 2004).

\subsubsection{Sulphur-bearing species}

\subsubsubsection{NS}

Our detection of NS is the first reported for IRAS~16293$-$2422.
In deriving its column density and abundance we have assumed a temperature
of 100~K, similar to that observed for the other hot core species.
Viti et al.\ (2001) predict that the abundance of NS is significantly
enhanced by shocks, and that a high NS/CS abundance ratio is indicative
of the evaporation of grain mantles with subsequent chemical processing
over a few $\times 10^4$~yr.  Comparison with the CS abundance derived
by Blake et al.\ (1994) gives NS/CS $\sim 0.6$ for IRAS~16293$-$2422,
similar to the Viti et al.\ models that require the grain mantles to have
evaporated early on, enabling significant subsequent chemical processing.
However, whether the evaporation is caused by sputtering in shocks or
by thermal evaporation is less easily determined from the models using
just the NS/CS ratio.

\subsubsubsection{H$_{\it 2}$S}

\begin{figure}
$$\includegraphics[scale=0.65,angle=-90]{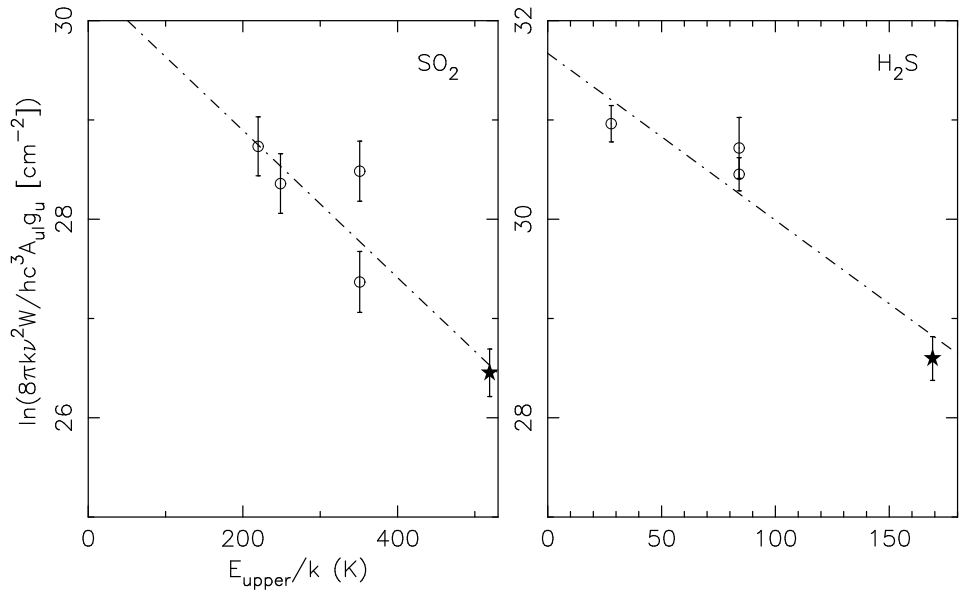}$$
\caption{Rotation diagrams for the sulphur-bearing species detected;
data from this work are marked as filled stars.  Supplementary SO$_2$
data come from Blake et al.\ (1994), and the dot-dash line is for $T_{\rm
rot} = 135$~K\@.  Supplementary H$_2$S data come from Blake et al.\
and Wakelam et al.\ (2004b), and the dot-dash line is for $T_{\rm rot}
= 60$~K.  \label{rot_sulphur}}
\end{figure}

H$_2$S emission is associated with both sources A and B, with the emission
at B having a much narrower line width, as has been observed in some
other species (e.g., Kuan et al.\ 2004; also see the H$_2$S profile
insets in Fig.~\ref{sulphur}).  Its emission is somewhat extended, and
the rotation diagram gives $T_{\rm rot} = 60$~K (Fig.~\ref{rot_sulphur}).
However, the observed abundance, $\sim 9 \times 10^{-8}$, is significantly
higher than that predicted by the Doty et al.\ model, for which $\sim 3
\times 10^{-9}$ is obtained at $t = 3 \times 10^3$~yr, and $\la 10^{-11}$
for $t = 3 \times 10^4$~yr.  However, our abundance estimate for H$_{\it
2}$S is in very good agreement with that predicted by Sch\"oier et al.\
(2002).  The narrower line width of H$_{\it 2}$S toward source B suggests
that the column density and abundance toward this source may be lower
than the composite value given in Table~\ref{abundances}.

\subsubsubsection{SO and SO$_{\it 2}$}

SO and SO$_2$ are formed in the gas phase, but may also be present
in grain mantles.  The SO$_2$ emission reported here originates from
the highest energy SO$_2$ transition reported so far in the literature
for IRAS~16293$-$2422 ($E_{\rm upper} = 519.1$~K).  The emission lies
significantly above the best fit to the rotation diagram presented by
Blake et al.\ (1994) of $T_{\rm rot} = 95$~K, and results in considerable
curvature in that rotation diagram indicative of the presence of
temperature gradients.  Using only those lines with $E_{\rm upper}
> 200$~K, we find $T_{\rm rot} \sim 135$~K for the compact emission
(Fig.~\ref{rot_sulphur}), and have used this in deriving the column
density and abundances for both SO$_2$ and $^{34}$SO$_2$.  The resulting
isotopic ratio is $N$(SO$_2$)/$N$($^{34}$SO$_2$) = 17, similar to that
derived for S/$^{34}$S from CS in the interstellar medium, of $\sim 22$
(Wilson \& Rood 1994).  The measured abundance for SO$_2$, $\sim 1 \times
10^{-7}$, is an order of magnitude higher than that predicted by Doty
et al.\ (2004) of $1 \times 10^{-8}$, but is in good agreement with the
analysis of Sch\"oier et al.\ (2002).

The SO line is the strongest in the entire spectrum.  Its peak line
brightness temperature is 46~K, indicating significant optical depth.
Indeed, Blake et al.\ (1994) derive their rotation temperature of 80~K
from the $^{34}$SO isotopomer instead, because of high optical depth in
the transitions of SO they observed.  The measured abundance, $5 \times
10^{-7}$, is two orders of magnitude higher than that predicted by Doty
et al.\ (2004), of $\sim 4 \times 10^{-9}$. The SO abundance in the
Sch\"oier et al.\ (2002) model is within a factor of two of our estimate.

\subsubsubsection{Si$^{\it 34}$S}

We also report the possible detection of Si$^{34}$S, although it is
blended with both A- and E-type methyl formate emission, and should at
present be regarded as tentative.

\subsubsection{Deuterated species} 

The envelope surrounding IRAS~16293$-$2422 exhibits remarkably high
abundances of deuterated species.  Bright, extended D$_2$CO has been
observed (Ceccarelli et al.\ 1998, 2001; Loinard et al.\ 2000), and even
triply-deuterated methanol, CD$_3$OH (Parise et al.\ 2004a).  Typical
fractionations are 5--20\%, depending on the species (e.g., Roberts
et al.\ 2002; Parise et al.\ 2004a).  High deuterium fractionation is
thought to be the result of grain surface chemistry under low temperature
conditions, and the deuterated molecules are observed when they are
subsequently desorbed back into the gas phase (Ceccarelli et al.\ 2001).

\subsubsubsection{DNO}

Here we report the probable detection of deuterated HNO in absorption
against source A\@.  DNO has the best match in frequency for the
absorption feature at 300955~MHz, but there are no other reports of
either HNO or DNO observations in the literature for IRAS~16293$-$2422.
For other sources, however, Snyder et al.\ (1993) and Ziurys, Hollis, \&
Snyder (1994) suggest a NO/HNO abundance ratio of $\sim 100$--800, and
van Dishoeck et al.\ (1995) report an upper limit for a NO line towards
IRAS~16293$-$2422.  Unfortunately this upper limit is for a high energy
transition, and so does not give a very stringent upper limit on the
NO column density, $N$(NO) $< 1 \times 10^{19}$~cm$^{-2}$ (2$\sigma$).
Both NO and HNO are expected to be associated with the cooler, outer
envelope in IRAS~16293$-$2422 (see the predicted column densities of
Doty et al.\ 2004), and the energy of the 300955~MHz transitions are
relatively low, consistent with observing this line in absorption
and resolving out much of the extended emission from the envelope.
The optical depth of the absorption feature is $\approx 0.45$, which for
an excitation temperature of 15--20~K implies a column density $N$(DNO)
$\sim 3 \times 10^{15}$~cm$^{-2}$.  This is consistent with the NO upper
limit if NO/HNO $\sim 300$, and DNO/HNO $\sim 0.1$.

\subsubsubsection{HDO}

\begin{figure}
$$\includegraphics[scale=0.65,angle=-90]{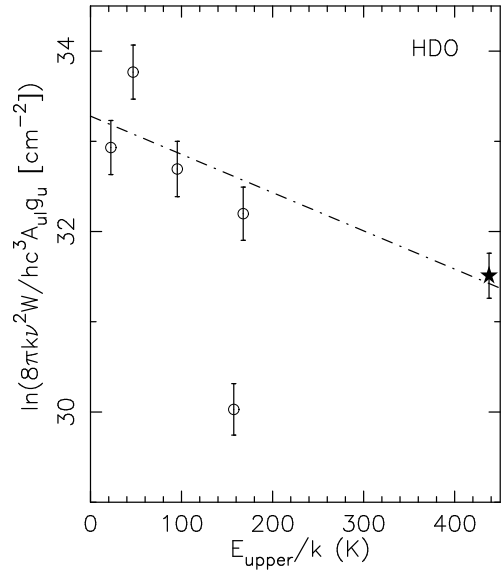}$$
\caption{Rotation diagram for HDO; data from this work are marked as
filled stars.  Supplementary data are from van Dishoeck et al.\ (1995),
Stark et al.\ (2004), and Parise et al.\ (2004b), and the dot-dash line
is for $T_{\rm rot} = 236$~K\@.  The low point at $E_{\rm upper}/k =
157$~K is not included in the fit, for the reasons described in the text.
\label{rot_deuterated}}
\end{figure}

We also detect a relatively high-energy transition of deuterated
water, HDO\@.  Parise et al.\ (2004b) have recently performed an
analysis of the HDO abundance in different components of the envelope
of IRAS~16293$-$2422, finding a considerably higher abundance in the
inner region, where the ices have evaporated from the dust grains,
compared with the outer envelope.  Fig.~\ref{deuterated} shows that the
emission region for the high-energy transition is indeed very compact.
In Fig.~\ref{rot_deuterated} we show the rotation diagram for HDO,
including other data from the literature.  The one point that lies
significantly away from the best-fit line for $T_{\rm rot} = 236$~K is
a measurement from Parise et al.\ (2004b) obtained with a 10$''$ beam
pointed at source B\@.  Since the HDO emission is in fact associated
with source A, the emission was probably close to the half-power point
of the beam (or worse, if the pointing was off by only a few arcsec).
We have therefore not used this point in the fit.  The derived abundance
is about a factor of 4 lower than that obtained by Parise et al.

\subsubsection{Unidentified lines}

Emission from a number of unidentified lines is detected, primarily
associated with source A (Fig.~\ref{deuterated}).  The 300899~MHz line is
particularly bright, and has immediately to its redshifted side a strong
absorption toward source A\@.  It is possible that these originate from
completely different lines, but in the absence of further information
they have both been designated as 300899~MHz here.

\subsubsection{Comparison with the physical-chemical model for
IRAS~16293$-$2422}
\label{chem_compare}

In general the physical-chemical model of Doty et al.\ (2004) on small
size scales matches well the observed abundances of first generation
precursors of organic species, i.e., H$_2$CO and CH$_3$OH, but the
abundance of second generation molecules, CH$_3$OCH$_3$ and HC$_3$N, are
underestimated by about an order of magnitude.  For the sulphur-bearing
species all abundances are underestimated by the physical-chemical model
of Doty et al.\ (2004) by one to two orders of magnitude.  Overall the
agreement with the empirical model of Sch\"oier et al. (2002) on small
size scales is much better.  These results suggest that the chemistry
of first generation species is well understood, but that of second
generation organic molecules and sulphur bearing species are not (see
also the discussion of Doty et al.\ 2004).

There are two possible explanations for the low abundances of sulphur
bearing species in the Doty et al.\ (2004) model.  First, Wakelam et
al.\ (2004a) have recently shown that the abundances of sulphur-bearing
species such as SO, SO$_2$, and H$_2$S in the hot core phase are strongly
dependent on the assumed form and abundance of sulphur-bearing species
on the dust grains before evaporation.  Moreover, a number of recent
experiments suggest that H$_2$S cannot be the sole carrier of sulphur on
dust grains (see, e.g., Wakelam et al.\ 2004a; van der Tak et al.\ 2003,
and references therein).  Indeed, Wakelam et al.\ (2004a) find the best
agreement with observations of both Orion and IRAS~16293$-$2422 for the
following ice composition: H$_2$S, $10^{-7}$; OCS, $10^{-7}$; and S,
$3\times 10^{-6}$ (atomic sulphur in molecular matrix).  In contrast,
Doty et al.\ (2004) assume that initially all of the sulphur is in the
form of H$_2$S.  Doty et al.\ (2002) report results from using a chemical
network analysis for the high mass hot core AFGL~2591 similar to that used
for their more recent study of IRAS~16293$-$2422.  In order to obtain
a good fit to the observed AFGL~2591 hot core SO$_2$ abundance, Doty et
al.\ (2002) adjusted the initial H$_2$S abundance until a good match was
achieved; the best fit was for an initial H$_2$S abundance of $1.6\times
10^{-6}$.  The same initial abundance of H$_2$S was assumed for the
Doty et al.\ (2004) IRAS~16293$-$2422 study, and indeed the single dish
SO$_2$ abundances for IRAS~16293$-$2422 and AFGL~2591 are quite similar
(Blake et al.\ 1994; van der Tak et al.\ 2003).  Likewise, the Doty et
al.\ SO$_2$ abundance is closer to the SMA result than for either SO or
H$_2$S, suggesting that the abundance of these latter two species are more
sensitive to the ice-phase sulphur assumptions.  Thus, the Doty et al.\
(2004) under-abundance of sulphur-bearing molecules may well be due the
their assumptions about the initial ice-phase sulphur carriers/abundances.

A second possibility is that shock chemistry (not included in the chemical
network of Doty et al.) has significantly affected the abundances of the
sulphur species in IRAS~16293$-$2422.  For example, Minh et al.\ (1990)
find that the abundances of SO, SO$_2$, and H$_2$S are enhanced by 1 to
2 orders of magnitude in the Orion Plateau region compared to average
hot core values (see, e.g., van der Tak et al.\ 2003; Hatchell et al.\
1998) and suggest shock chemistry is responsible.  Evidence for such a
shock at source A is discussed in Section~\ref{discussion}.

Figs.~\ref{rot_organic} to \ref{rot_deuterated} also demonstrate
the importance of interferometry at submillimeter wavelengths, using
instruments such as the SMA, to constrain molecular excitation conditions
from high-energy transitions.

\section{Discussion}
\label{discussion}

\subsection{Origin of the continuum emission}
\label{continuum_origin}

\subsubsection{Centimeter emission from source A}

The continuum spectrum for source A (Fig.~\ref{sed}) is well-fitted by
the combination of two separate power laws.  The emission from source
A includes a partially optically-thick free-free component at $\nu <
100$~GHz with $\alpha = 0.53\pm0.04$.  Such spectral indices can be
readily explained as originating in an ionized jet (Reynolds 1986).
For such jets the emission at each frequency arises from plasma
having $\tau_\nu \ga 1$ (where for free-free emission $\tau_\nu \propto
\nu^{-2.1}$), and spectral indices similar to that observed for source A
have been observed in several other sources, such as the driving source
of the HH1/2 system (Rodr\'\i guez et al.\ 1990) and Cep A HW2 (Garay
et al.\ 1996).  In the case of a fully ionized, isothermal jet with a
constant temperature and velocity, and where the transverse width of
the jet, $w$, has a radial dependence $w \propto r^\epsilon$, the value
of $\alpha$ is related to $\epsilon$ by $\alpha = 1.3 - 0.7/\epsilon$.
Further, the expected length of the observed radio jet has a frequency
dependence $\theta \propto \nu^{-0.7/\epsilon}$, and if observed with
sufficiently high spatial resolution to be able to separate the two
lobes of the jet, the distance between the lobes will also scale as
$\nu^{-0.7/\epsilon}$ (Rodr\'\i guez et al.\ 1990).  If we were to
assume that the centimeter radio emission from A1 and A2 is associated
with a wind or jet originating somewhere close to Aa, however, then the
expected frequency dependence of jet length and separation on frequency
is clearly not observed, since the separation is constant with frequency
(and with time) to within the measurement errors, at $0.35'' \pm 0.02''$.
We must therefore consider other possibilities.

\begin{figure}
$$\includegraphics[scale=0.65,angle=-90]{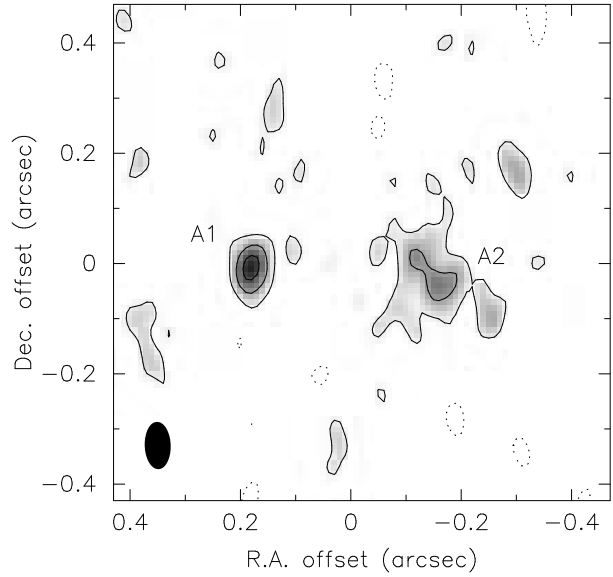}$$
\caption{43~GHz continuum image of sources A1 and A2.  The synthesized
beam is $86 \times 47$~mas at P.A. 2$^\circ$, shown at bottom left.
Contours are spaced at approximately 2$\sigma$ intervals of
0.25~mJy~beam$^{-1}$.  \label{a1_q}}
\end{figure}

The position of A2 relative to B has remained constant over the past 17
years, exhibiting no residual proper motion that might be expected if A1
and A2 were part of a binary system with $M_{\rm A1} \sim M_{\rm A2}$,
or if they were the oppositely-directed lobes of a precessing radio jet.
The centimeter radio emission from A2 therefore probably pinpoints the
location of a protostar.  Indeed, when observed at the highest resolution
available at 43~GHz, A2 appears to be bipolar with a position angle $\sim
45^\circ$, aligned with the large-scale northeast--southwest outflow
(Fig.~\ref{a1_q}).  Thus A2 is probably the driving source of this flow.

The source of the ionization in low luminosity protostars is generally
thought to be the result of shocks: either accretion shocks very close
to the protostar (e.g., Neufeld \& Hollenbach 1996; Shang et al.\ 2004),
or shocks in an otherwise neutral wind (Curiel, Cant\'o, \& Rodr\'\i
guez 1987; Ghavamian \& Hartigan 1998; Gonz\'alez \& Cant\'o 2002).
In the case of source A2, these shocks are probably internal shocks in a
jet, giving rise to the bipolar structure in Fig.~\ref{a1_q}.  The case
of A1 is more puzzling.  The separation A1$-$A2 remains remarkably
constant, even though their position angle changes dramatically.
This would suggest a physical association of these two sources, but
even though A1 and A2 have comparable radio flux densities, A2 does not
exhibit any reflex motion as a result of a gravitational interaction
--- the 3$\sigma$ upper limit to its motion relative to source B is
$4.5 (D/{\rm 160~pc})$~km~s$^{-1}$.  If the mass of A1 were instead
much lower than A2, and its motion relative to A2 is interpreted as
rotation, then the linear fit to the position angle as a function of
time shown in Fig.~\ref{pangle} would correspond to a rotation period of
$162 \pm 19$~yr.  For a (projected) separation of 0.35$''$ [$56(D/{\rm
160~pc})$~AU] this would imply a mass of $6.5(D/{\rm 160~pc})^3 M_\odot$
for the protostar at A2.  This mass is comparable to the mass of the
entire envelope, $\sim 5(D/{\rm 160~pc})^2 M_\odot$ (e.g., Sch\"oier
et al.\ 2002), and is inconsistent with the low total luminosity of
IRAS~16293$-$2422, $27(D/{\rm 160~pc})^2 L_\odot$ (Mundy et al.\ 1986).
Even assuming a distance of 120~pc we would find a mass for A2 of $2.7
M_\odot$, an envelope mass of $\sim 3 M_\odot$, and a luminosity of $15
L_\odot$, which remains inconsistent.  If A1 were in a highly elliptical
orbit and observed at periastron the mass of A2 could be lowered a little,
but A1 will be unbound for A2 masses lower than $3.3(D/{\rm 160~pc})^3
M_\odot$.  Furthermore, the probability of observing A1 at periastron for
a highly elliptical orbit is very low.  It is therefore very unlikely that
the motion of A1 relative to A2 represents a gravitationally bound system.

The fact that the position angle of A1 relative to A2 has traversed the
rather wide opening angle of the east--west outflow from IRAS~16293$-$2422
leads us instead to interpret A1 as the location of a shock interaction
between the precessing jet or collimated wind responsible for this flow
and nearby dense gas.  The close association of A1 with the (probable)
torsionally-excited methanol emission lends support to this picture.  The
origin of this flow is discussed further in Section~\ref{outflow_origin}.

The peak brightness temperatures ($T_{\rm B}$) for the radio emission
from source A2 are $\sim 300$~K at 8~GHz, $\sim 160$~K at 22~GHz,
and $\sim 100$~K at 43~GHz.  For A1 they are $\sim 420$~K at 8~GHz,
$\sim 190$~K, and $\sim 140$~K at 43~GHz.  Assuming the emission at each
frequency to be dominated by gas with $\tau \ga 1$ at $T = 10^4$~K, as
indicated by the centimeter spectral index, these brightness temperatures
can be used to estimate source solid angles from $\Omega_{\rm source} =
\Omega_{\rm beam} \times (T_{\rm B}/10^4$~K), which can then be used to
estimate the linear dimension of the emitting region.  The brightness
temperatures above therefore imply linear dimensions for the shocked
emission at A2 of $\sim 6 \times 10^{13}(D/{\rm 160~pc})$~cm at 8~GHz,
$\sim 2 \times 10^{13}(D/{\rm 160~pc})$~cm at 22~GHz, and $\sim 9 \times
10^{12}(D/{\rm 160~pc})$~cm at 43~GHz.  Similarly for A1, the linear
dimensions are $\sim 8 \times 10^{13}(D/{\rm 160~pc})$~cm at 8~GHz,
$\sim 3 \times 10^{13}(D/{\rm 160~pc})$~cm at 22~GHz, and $\sim 1 \times
10^{13}(D/{\rm 160~pc})$~cm at 43~GHz.  Individually, therefore, A1 and
A2 do exhibit the decrease in source size as a function of frequency
expected for partially optically-thick free-free emission.

\subsubsection{Centimeter emission from source B}

\begin{figure}
$$\includegraphics[scale=0.65,angle=-90]{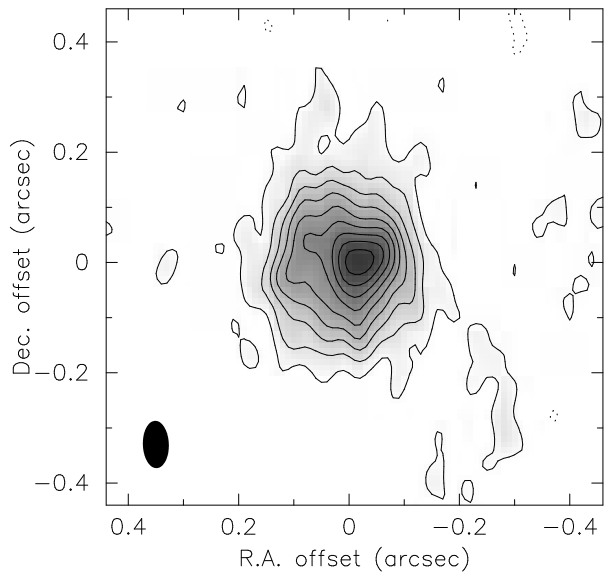}$$
\caption{43~GHz continuum image of source B\@.  The synthesized
beam and contours are as for Fig.~\ref{a1_q}.  \label{b_q}}
\end{figure}

Source B exhibits a centimeter radio spectrum consistent with being
optically thick and thermal.  The emission is much more centrally-peaked
than is the case for source A, and formal Gaussian fits to the emission
shown in Fig.~\ref{overlays} are for 15~GHz (1987.66) and 8~GHz (1989.05,
2002.41, and 2003.65) consistent with circular symmetry.  At 22~GHz the
emission has a position angle $\approx 22\pm31^\circ$, and at 8~GHz,
epoch 1994.30, the position angle is $\approx -10\pm18^\circ$, consistent
with that at 22~GHz, but both have large uncertainties.  The deconvolved
source size is $0.19'' \times 0.15''$ at 22~GHz, and $0.16'' \times
0.10''$ at 8~GHz.  At 43~GHz the emission is resolved and the lower
contours are approximately circular, with a peak slightly offset from
the centroid of the outer contours (Fig.~\ref{b_q}; see also Rodr\'\i
guez et al.\ 2005).  There are two possible origins for this emission:
optically-thick free-free emission from ionized gas, or optically-thick
dust emission.  The trend in deconvolved source size with frequency
is opposite to that expected for free-free emission, but is consistent
with dust.  The peak brightness temperatures observed are $\sim 160$~K
at 8~GHz in a beam $0.35'' \times 0.17''$, $\sim 350$~K at 22~GHz in a
beam $0.194'' \times 0.090''$ (see also Mundy et al.\ 1992), and $\sim
390$~K at 43~GHz in a beam $86 \times 47$~mas.  Thus temperatures of 390~K
would appear to occur at a physical radius $\sim 5 (D/{\rm 160~pc})$~AU,
350~K at $11(D/{\rm 160~pc})$~AU, and 160~K at $20(D/{\rm 160~pc})$~AU\@.
This suggests a temperature gradient consistent with internal heating
by a central protostar, although more detailed modelling is needed to
establish the radial temperature profile.

In order for the $\nu^2$ dust spectrum of source B to extend to
frequencies as low as 5~GHz either the dust column density must be
very high, or the dust grains must be large, of order a few centimeter
(e.g., Miyake \& Nakagawa 1993).  We cannot rule out a contribution to
the 5~GHz emission from ionized gas, but consider the 8~GHz emission to
be due to the same emission mechanism as at 15~GHz, 22~GHz, and 43~GHz.
Given the high brightness temperature observed for the radio emission it
is most likely that it is optically thick dust emission, although there
may be large particles having opacities $\kappa_\nu \propto \nu^\beta$,
with $\beta \sim 0$, as well.  To estimate the mass of dust and gas for
source B we assume a dust opacity suitable for particle densities $n_{\rm
H} \sim 10^8$~cm$^{-3}$, for coagulated particles with no ice mantles,
from Table 1, column 3, of Ossenkopf \& Henning (1994).  These authors
give $\kappa = 5.86$~cm$^2$~g$^{-1}$ at $\nu = 230$~GHz, and from the
lowest frequencies given in their Table~1, we find $\beta = 1.1$, which
we use to extrapolate to lower frequencies.  Including a gas to dust
ratio, $M_{\rm g}/M_{\rm d} = 100$, we find $\kappa_{\rm 8~GHz} \sim 1.5
\times 10^{-3}$~cm$^2$~g$^{-1}$ and $\kappa_{\rm 43~GHz} \sim 9.3 \times
10^{-3}$~cm$^2$~g$^{-1}$, where $\kappa_{\rm 8~GHz}$ and $\kappa_{\rm
43~GHz}$ are now per gram of gas plus dust.  Thus in order for the 8~GHz
emission to be optically-thick ($\tau \sim \kappa N_{\rm H} m_{\rm H}
\mu \sim 1$, where $N_{\rm H}$ is the column density of hydrogen atoms,
$m_{\rm H}$ is the mass of a hydrogen atom, and $\mu \approx 1.36$
is the ratio of total gas mass to hydrogen mass: Hildebrand 1983),
a particle column density $N_{\rm H} \sim 3 \times 10^{26}$~cm$^{-2}$
is needed.  For the 43~GHz emission to be optically-thick, $N_{\rm H}
\sim 5 \times 10^{25}$~cm$^{-2}$ is needed.

The deconvolved source size at 8~GHz corresponds to a radius $r \sim
11(D/{\rm 160~pc})$~AU, implying a total mass for a uniform source of $M
= \pi r^2 N_{\rm H} m_{\rm H} \mu \sim 0.03(D/{\rm 160~pc})^2 M_\odot$.
If the column density has a steeper radial dependence, for example
$N_{\rm H} \propto r^{-3/2}$ as expected for an accretion disk, then the
total mass may somewhat higher, $\sim 0.12(D/{\rm 160~pc})^2 M_\odot$.
At 43~GHz the emission extends out to radii of 0.2$''$ [$32(D/{\rm
160~pc})$~AU], with the lowest contour in Fig.~\ref{b_q} corresponding
to a brightness temperature of 40~K\@.  A similar calculation for the
43~GHz emission produces a total mass for a uniform source of $\sim
0.04(D/{\rm 160~pc})^2 M_\odot$ at 43~GHz, and $\sim 0.16(D/{\rm
160~pc})^2 M_\odot$ for a column density proportional to $r^{-3/2}$.
Now, for a disk with surface density $\Sigma \propto r^{-p}$, $p > 0$, it
can easily be shown that the radius at which $\tau = 1$ is proportional
to $\nu^{\beta/p}$.  The size of the 8~GHz emission compared with that
at 43~GHz would suggest $\beta/p \sim 0.6$, similar to that expected for
a disk model with $\beta \sim 1$ and $p \sim 3/2$.  Note that although
the emissivity of dust is highly uncertain at centimeter wavelengths,
these masses are nevertheless reasonable, and can account for the
observed radio emission.  We do not find a mass as high as that derived
using a similar surface density distribution by Rodr\'\i guez et al.\
(2005) of 0.5--0.7~$(D/{\rm 160~pc})^2 M_\odot$, which we attribute to
a different assumed dust opacity.

We now take the peak brightness temperatures per beam, and the physical
size scale of the corresponding beam at the source, to estimate the
contribution of source B to the total luminosity for IRAS~16293$-$2422
derived from the spectral energy distribution of $27(D/{\rm 160~pc})^2
L_\odot$ (Mundy et al.\ 1986).  In an exact 1D radiative transfer solution
for dusty protostellar envelopes, for the case where the radiation field
is optically-thick in the far-infrared, Preibisch, Sonnhalter, \& Yorke
(1995) show that in the temperature range $\sim 150$--400~K the radiation
temperature is approximately 25\% higher than that expected from simply
assuming $L = 4\pi r^2 \sigma T^4$.  We therefore estimate the luminosity
of source B from the observed brightness temperatures using $L = 4\pi r^2
\sigma (0.8T)^4$, and find that $T = 160$~K at radius $r \sim 20(D/{\rm
160~pc})$~AU gives $L \sim 4.4(D/{\rm 160~pc})^2 L_\odot$; $T = 350$~K
at $r \sim 11(D/{\rm 160~pc})$~AU gives $L \sim 31(D/{\rm 160~pc})^2
L_\odot$; and $T = 390$~K at $r \sim 5(D/{\rm 160~pc})$~AU gives $L \sim
10(D/{\rm 160~pc})^2 L_\odot$.  Although these numbers are only rough
estimates, the brightness temperature derived from the 22~GHz emission
in particular indicates that source B is responsible for a least half
the total luminosity, and indeed may suggest that the radiation field
probably has to be anisotropic.  One possibility that can reconcile this
luminosity estimate and the observed morphology of the radio source
is that the emission arises from a face-on disk.  We can therefore
conclusively rule out the possibility that source B is a low-luminosity,
young T Tauri star, as has been suggested by Stark et al.\ (2004).

\subsubsection{Submillimeter emission from A and B}

At $\nu > 100$~GHz the spectra of sources A and B are consistent with
optically-thin dust emission with $\beta \ga 0.8$ for source A, and $\beta
\ga 0.4$ for source B\@.  As described above in Section~\ref{results} the
different $u$-$v$ coverage for each of the millimeter and submillimeter
data points in Fig.~\ref{sed} make the derivation of the spectral
index rather more uncertain than the errors quoted in the figure.
However, the overall spectrum of the entire region is well-fitted by a
single-temperature greybody with $T_{\rm dust} = 40 \pm 1$~K and $\beta =
1.6$ (Stark et al.\ 2004), which for optically-thin emission gives a 300
to 110~GHz spectral index of $\alpha = 3.48$, somewhat steeper than that
observed for the individual sources in Fig.~\ref{sed}.  The brightness
temperatures of the submillimeter emission obtained from the image
shown in Fig.~\ref{cont_su} restored with a normal CLEAN beam of $1.23''
\times 0.64''$ at P.A. $10^\circ$ (instead of a circular 0.4$''$ beam)
are 21~K for source A, and 38~K for source B\@.  Since the temperature
of dust in the envelope on these scales is probably less than 100~K, the
submillimeter dust emission from the envelope clearly has significant
optical depth at $\sim 150$~AU, that may account for the shallower
spectral index of the emission detected by the interferometers compared
with the overall envelope.

A striking feature of the dust continuum emission from source A is that
even on subarcsecond size scales there is little evidence of flattened
structures that one might call a ``disk.''  For the case of the collapse
of a slowly rotating, singular isothermal sphere, the centrifugal
radius is given by $R = \Omega^2G^3M^3/16a^8$, where $\Omega$ is the
angular velocity of the cloud, $M$ is the central mass, and $a$ is the
isothermal sound speed (Terebey, Shu, \& Cassen 1984; hereafter TSC).
The kinematics of the surrounding cloud core have been examined by Zhou
(1995) and Narayanan et al.\ (1998) in terms of the TSC model, for
which the best fits give $\Omega \approx 4 \times 10^{-13}$~s$^{-1}$,
and $a \approx 0.5$~km~s$^{-1}$.  For a 1~$M_\odot$ central object this
gives $R \sim 40$~AU ($\sim 0.3''$).  Note that there is a very steep
dependence of $R$ on $a$, and the cloud is not isothermal.  Nevertheless,
this calculation may explain why, even with the resolution available
with the SMA, the dust emission does not look ``disk-like''; any disk
remains below the resolution limit of the current submillimeter data.

The location of Aa almost exactly between the radio sources A1 and
A2 implies either that there is a protostar at this position, or that
the flux ratio of any dust emission associated with A1 and A2 is close
to unity.  The interpretation of A1 as shock-ionized gas resulting from
a wind or jet impacting dense gas would suggest that A1 should also be
the location of significant dust emission; if so, a model where A1 and A2
are both sources of submillimeter emission, with roughly equal fluxes,
would seem to explain the data best.  Higher resolution submillimeter
observations are clearly needed to establish whether there is really a
separate source at the position of Aa.

In order to obtain lower limits on the masses of the compact components
listed in Table~\ref{compact} we assume a dust opacity $\kappa_{\rm
300~GHz} \sim 1.5$~cm$^2$~g$^{-1}$ appropriate for dust grains with ice
mantles from Ossenkopf \& Henning (1994, Table 1), $M_{\rm g}/M_{\rm d}
= 100$, and $T_{\rm dust} = 40$~K\@.  The particle column density needed
for $\tau \sim 1$ is then $N_{\rm H} \sim 3 \times 10^{25}$~cm$^{-2}$,
and the corresponding masses for a uniform source are $M_{\rm Aa} \sim
0.14(D/{\rm 160~pc})^2 M_\odot$, $M_{\rm Ab} \sim 0.04(D/{\rm 160~pc})^2
M_\odot$, and $M_{\rm B} \sim 0.23(D/{\rm 160~pc})^2 M_\odot$.  Note that
extrapolating the centimeter wavelength emission from source B to 300~GHz
(Fig.~\ref{sed}) suggests a contribution to the 300~GHz flux from the
optically-thick disk of $\sim 1.4$~Jy, which is $\sim 50$\% of the compact
emission from Table~\ref{compact}.  Assuming $p = 3/2$ so that the mass
of the disk is $\sim 0.16(D/{\rm 160~pc})^2 M_\odot$, the contribution
from the envelope is then $\sim 0.11(D/{\rm 160~pc})^2 M_\odot$, and
the total is $\sim 0.27(D/{\rm 160~pc})^2 M_\odot$, similar to the value
obtained for a uniform source with a lower overall value of $\kappa$.

\subsection{Relationship between the radio sources and the large-scale
outflows}
\label{outflow_origin}

At least two outflows have been identified originating from the vicinity
of IRAS~16293$-$2422, one oriented NE--SW with its redshifted lobe to
the northeast, and the other approximately E--W with its redshifted lobe
to the west (Mizuno et al.\ 1990; Stark et al.\ 2004).  Mizuno et al.\
(1990) also detect a low-velocity, unipolar blueshifted lobe extending 10
arcmin to the east, which Stark et al.\ suggest is just an extension of
the E--W flow.  The redshifted counterpart of this blue lobe has probably
already broken out of the molecular cloud to the west.  The NE--SW
flow is well-collimated, and centered on source A\@.  Estimates of
the inclination to the plane of the sky for this flow range from
30--45$^\circ$ (Hirano et al.\ 2001) to 65$^\circ$ (Stark et al.\ 2004).
There is, however, considerable low-velocity red and blueshifted emission
spatially coincident with the blue and redshifted lobes of this flow
respectively in the maps of Stark et al., suggesting that an inclination
closer to the plane of the sky than 65$^\circ$ may be most appropriate.
We have identified the radio source A2 as the source of this flow, based
on its apparent bipolar morphology aligned with the molecular outflow.

The red and blue peaks of the E--W flow are centered somewhat to the
north of source A, but the emission is messy, and at low velocities,
close to IRAS~16293$-$2422, the contours in the maps of Stark et al.\
(2004) are in fact also centered close to source A\@.  The E--W flow
exhibits a larger opening angle than the NE--SW flow.  The orientation
of the E--W flow is such that its interaction with I16293E, a nearby and
possibly starless dark cloud 1.5$'$ to the east of IRAS~16293$-$2422,
has disrupted the flow, as it streams around the edges of this cloud.
These features led Stark et al.\ to suggest that the E--W flow is a fossil
flow driven by source B\@.  Such an interpretation would imply that
source B is older than source A, and that source B is a low-luminosity
pre-main sequence star.  This would seem to be in contradiction with
the observations of source B which indicate some of the same features as
deeply-embedded class 0 protostars, such as the presence of some of the
same hot core molecules as are observed for source A, and with the high
luminosity implied by the brightness temperatures derived from the radio
emission in Section~\ref{continuum_origin}.  Furthermore, the narrow
linewidths associated with source B would suggest instead that source
B is younger than source A (a conclusion also reached by Wootten 1989).

Copious water masers are associated with various shocks at sources
Aa/A1/A2, and exhibit expansion proper motions of $\sim 65(D/{\rm
160~pc})$~km~s$^{-1}$ in the plane of the sky (Wootten et al.\ 1999),
considerably higher than the radial velocities observed in the molecular
component of the outflows on larger scales, of 10--20~km~s$^{-1}$.
Furthermore, the motion of A1, interpreted above as tracing the
location of the shock interaction of dense gas with a precessing jet,
makes the origin of this jet a good candidate for explaining some of
the features of the E--W outflow, such as its wide opening angle, its
apparent recollimation on the far side of I16293E (Stark et al.\ 2004),
and the collimated unipolar blueshifted flow (Mizuno et al.\ 1990):
the apparent recollimation and the unipolar flow would be interpreted
as directions of lower density that have been evacuated during previous
epochs of the jet passing in these directions.

So, where is the driving source of this precessing jet?  The rate at
which the jet is precessing may provide a clue.  The precession of jets
has been inferred for some time now to explain S-shaped optical jets and
Herbig-Haro flows (e.g., Bally \& Devine 1994; Schwartz \& Greene 1999),
and the presence of multiple bow shocks in molecular outflows such as
RNO~43 and L1157 (Bence, Richer, \& Padman 1996; Gueth et al.\ 1997).
However, such a dramatic change in position angle over a relatively short
time baseline, such as that presented here for IRAS~16293$-$2422, has
not been observed directly before.  Various authors have investigated the
possibility that the tidal interaction between a star with a circumstellar
disk and a binary companion in an orbit misaligned with the disk might
cause jet precession, assuming the jet to originate in the inner regions
of the disk (e.g., Terquem et al.\ 1999; Bate et al.\ 2000).  In these
scenarios, the disk precesses about the orbital axis with a period $P_{\rm
prec} \sim 20 P_{\rm orb}$, where $P_{\rm orb}$ is the orbital period.
On top of this precession, the disk might wobble in response to the torque
exerted on the disk in the direction aligning it with the binary orbit,
with a wobble period $\sim P_{\rm orb}/2$ (Bate et al.\ 2000).

If the slight curvature in the observed rate of change of position angle
as a function of time in Fig.~\ref{pangle} is interpreted as precession,
the precession (or wobble) period of the jet is $\la 100$~yr.  Assuming
the wobble to be induced by a binary system with period $\la 200$~yr the
binary separation would have to be $\la 22$~AU (0.14$''$) for the wobble
described above, or $\la 2$~AU for precession, assuming a total stellar
mass of 1~$M_\odot$.  Since we have already established that the radio
emission from A2 is probably associated with a protostar, we therefore
look to a second source in the vicinity of A2, in an inclined orbit to
A2, as the origin of the precessing jet that drives the E--W molecular
outflow.  We note that, although Aa is not yet established as a separate
source rather than the centroid of submillimeter emission from A1 and
A2, the separation between A2 and Aa is similar to the binary separation
required if Aa were the origin of the E--W outflow and the precessing jet.
The new submillimeter source Ab has a projected separation from Aa of
0.64$''$, or 102~AU\@.  Assuming a circular orbit this would result in
an orbital period $\sim 10^3$~yr, and a wobble period $\sim 500$~yr,
clearly too long to account for the change in position angle of the
centimeter radio emission.

The total CO momentum rate in the E--W flow is $\sim 4 \times
10^{-4}$~$M_\odot$~km~s$^{-1}$~yr$^{-1}$ (including the unipolar
blueshifted lobs), and that of the NE--SW flow is $\sim 1 \times
10^{-4}$~$M_\odot$~km~s$^{-1}$~yr$^{-1}$ (Mizuno et al.\ 1990).  While
there is large scatter in the observed correlations between 5~GHz radio
emission and the CO momentum rate (Cabrit \& Bertout 1992; Anglada 1995),
these momentum rates are certainly sufficient to account for the radio
emission in terms of shock ionization.

\begin{figure}
$$\includegraphics[scale=0.65,angle=-90]{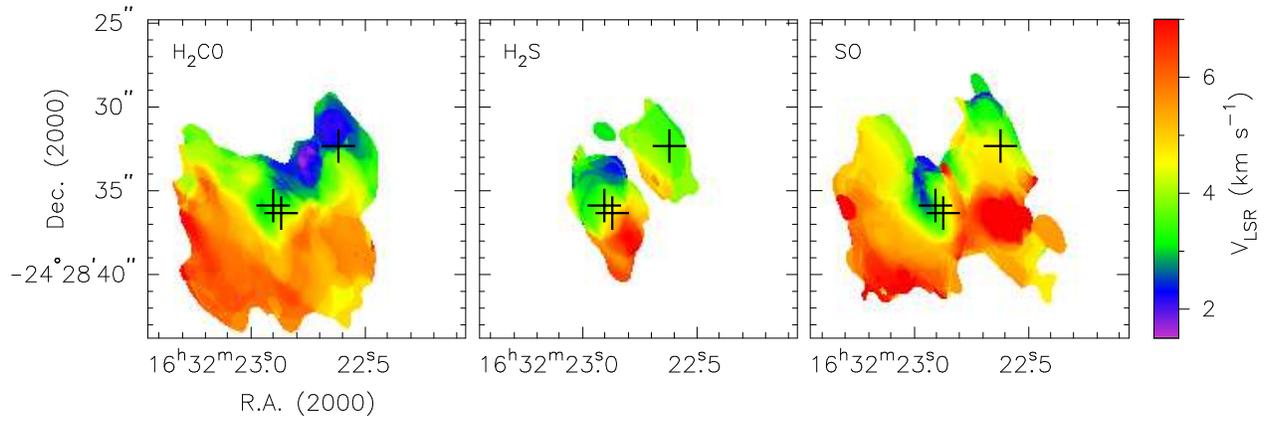}$$
\caption{Intensity-weighted mean radial velocity for H$_2$CO, H$_2$S,
and SO emission.  The positions for sources Aa, Ab, and B from
Table~\ref{compact} are marked as crosses.  \label{mom1_images}}
\end{figure}

The passage of a jet with variable directions and inclinations to the
plane of the sky also helps to explain the kinematics of the molecular
gas on small scales close to source A\@.  Fig.~\ref{mom1_images}
shows the intensity-weighted mean velocity (first moment) images of
the formaldehyde, H$_2$S, and SO emission.  There is a clear velocity
gradient that lies perpendicular to the NE--SW outflow, along a line
connecting sources A and B, that has been investigated in a number of
single-dish and interferometric studies and interpreted as rotation about
an axis approximately aligned with this outflow (Mundy et al.\ 1986;
Menten et al.\ 1987; Walker, Carlstrom, \& Bieging 1993).  However, the
kinematics are complicated by emission from the outflows themselves, made
all the more confusing because the dominant velocity gradients along the
outflow directions are actually reversed from those of the large-scale,
high-velocity CO emission.

Fig.~\ref{mom1_images} illustrates widespread redshifted formaldehyde
emission to the south of source A, while the blueshifted emission lies
predominantly to the northwest of A, along the axis joining A and B\@.
This geometry is almost exactly the opposite of the kinematics in the
CO outflows.  For H$_2$S the emission is more localized around sources
A and B, with B having a narrow linewidth, $\sim 1.6$~km~s$^{-1}$,
compared with $\sim 6$~km~s$^{-1}$ for source A (Fig.~\ref{sulphur}).
Across source B there is a shallow velocity gradient running NW--SE,
while the velocity gradient associated with source A runs along the
axis of the NE--SW outflow, but in the opposite sense from the higher
velocity CO\@.  The SO first moment image is similar to the H$_2$CO,
but extends to higher velocities because of the higher signal-to-noise
ratio in the stronger line.  Two distinct axes can be defined for the
principal kinematic features, one along the line joining sources Aa
and Ab at P.A. 45$^\circ$, aligned with the NE--SW outflow, and one
perpendicular to this at P.A. 135$^\circ$, along the line joining source
A and source B, and aligned with the velocity gradient associated with
the proposed rotation.

\begin{figure}
$$\includegraphics[scale=0.7,angle=-90]{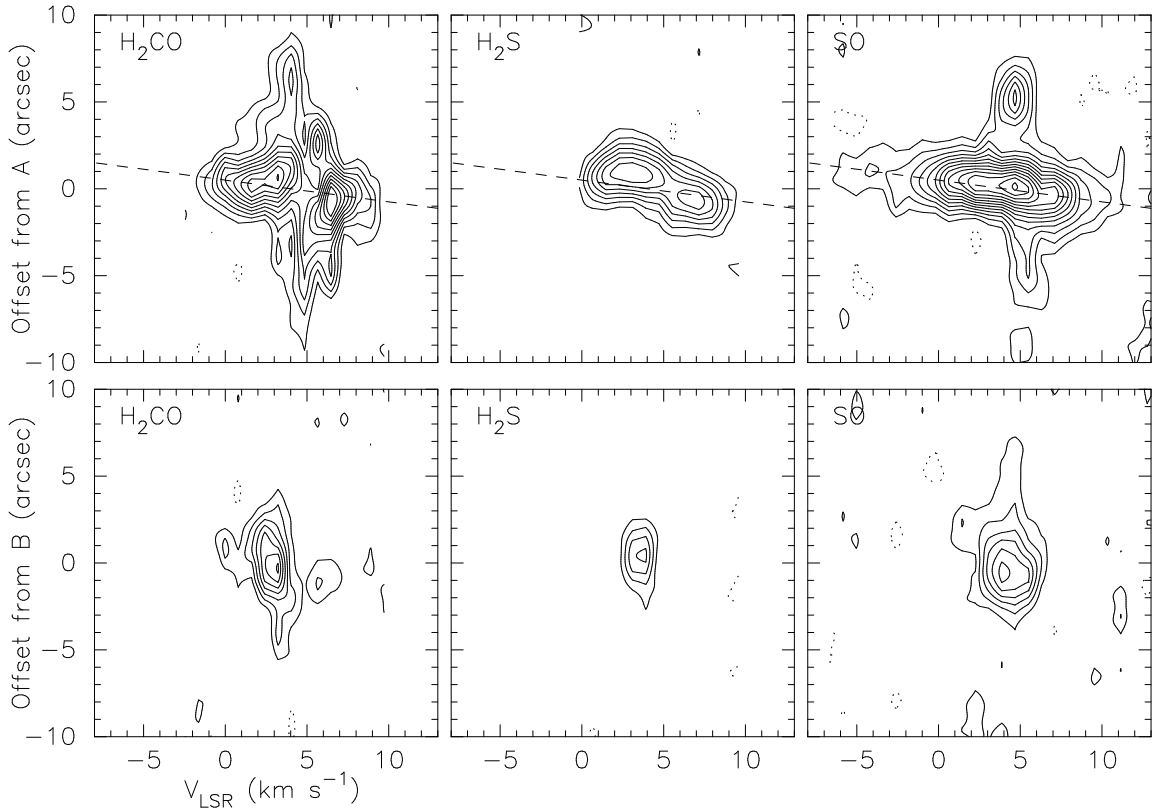}$$
\caption{Position-velocity diagrams for cuts at P.A. = 45$^\circ$ through
source A (top panels) and source B (bottom panels).  Contours are linearly
spaced at 0.65~Jy~beam$^{-1}$.  The dashed line in the panels for source A
shows the velocity gradient passing through the position of the continuum
peak at the systemic $V_{\rm LSR} = 4$~km~s$^{-1}$.  \label{pv_outflow}}
\end{figure}

\begin{figure}
$$\includegraphics[scale=0.7,angle=-90]{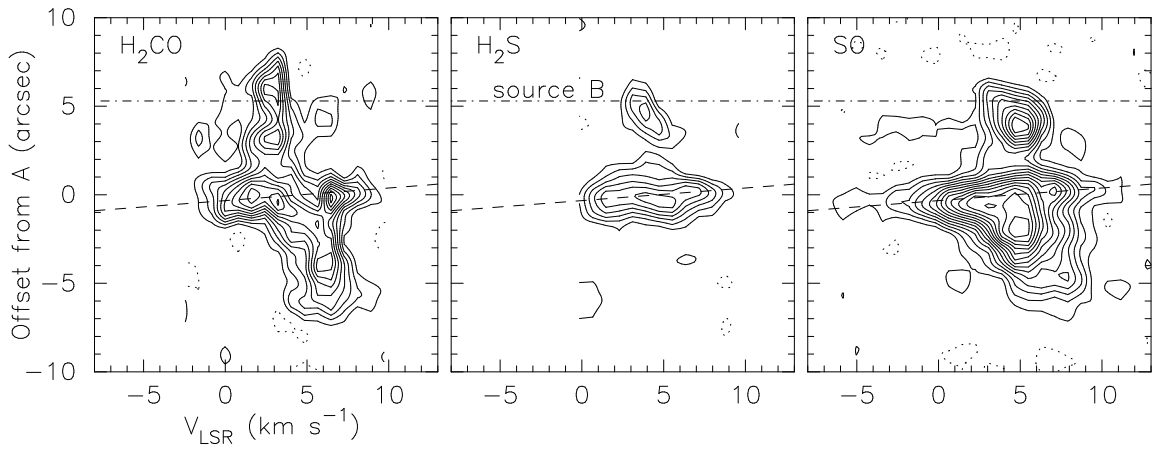}$$
\caption{Position-velocity diagrams for a cut at P.A. = 135$^\circ$ through
a line joining sources A and B\@.  Contours are as in Fig.~\ref{pv_outflow}.
The dashed line denotes the kinematic feature described in the text
having the velocity gradient with the same sense as the large-scale
CO outflow.  This kinematic feature passes through the position of the
continuum peak at source A and the systemic $V_{\rm LSR} = 4$~km~s$^{-1}$.
The dot-dash line shows the position of the continuum peak at source
B.  \label{pv_rotation}}
\end{figure}

Figs.~\ref{pv_outflow} and \ref{pv_rotation} show position-velocity
(P--V) diagrams for cuts along P.A. 45$^\circ$ aligned with the NE--SW
outflow for sources A and B, and for a cut at P.A. 135$^\circ$ along
a line connecting sources A and B\@.  The general agreement between
the overall structure observed in the two species exhibiting extended
emission, H$_2$CO and SO, is striking.  There is, however, one noticeable
difference: the formaldehyde is lacking emission close to the systemic
velocity of the cloud, $\approx 4$~km~s$^{-1}$, probably due to a
combination of self-absorption by the cooler, outer envelope at these
velocities, and the fact that the interferometer filters out much of the
extended emission at the cloud velocity.  Thus the apparent difference
in $V_{\rm LSR}$ for sources A and B, with B exhibiting $V_{\rm LSR}
\sim 3$~km~s$^{-1}$ for H$_2$CO, is probably exacerbated by this problem.
Indeed, for the H$_2$S and SO emission, which have much lower abundances
in the outer envelope and so do not exhibit such self-absorption, the
velocities of A and B are more similar.  The velocity gradient along
the axis connecting A and B no longer looks like rotation now that it
is observed at such high spatial resolution.  Rather, it is probably a
combination of the spatial filtering and self-absorption at the cloud
systemic velocity, and the enhancement of molecular emission in shocked
clumps as the outflow interacts with the envelope.

Although overall the sense of the red and blueshifted emission from the
high density tracers is in the opposite sense to that of the outflows,
the cut along axis A--B, Fig.~\ref{pv_rotation}, does reveal the presence
of a spatially compact, kinematic feature at source A (marked as a dashed
line in that figure), with velocities extending $\pm 6$~km~s$^{-1}$ from
the systemic velocity within $\pm 1''$ of the continuum source, and with
the same velocity sense as the large-scale CO outflows.  This feature
is masked by the bright, extended emission in the first moment maps,
but does lend support to the idea that the more energetic, E--W flow,
is driven by source A\@.  However, none of the P--V diagrams exhibit the
``typical'' outflow signatures observed in, e.g., CO emission, because
of the confusing effects of abundance enhancements in local shocks.

Fig.~\ref{pv_outflow} shows that at P.A. 45$^\circ$, along the direction
of the NE--SW outflow, both the H$_2$CO and SO emission from source
A is clumpy, and is extended particularly at velocities close to the
systemic velocity of the cloud, 4~km~s$^{-1}$.  The velocity gradient
of redshifted emission toward the southwest (negative position offsets)
to blueshifted in the northeast is predominantly within $\pm 1''$ of
the continuum peak, and the linear feature marked by the dashed line is
typical of the P--V diagrams resulting from rotating protostellar disks
(Richer \& Padman 1991).  If interpreted as rotation, the observed
velocity gradient would correspond to a total central mass for source
A of $\sim 1.5$--2~$M_\odot$.

The velocity gradient at source B is harder to interpret, because only the
H$_2$CO emission peaks close to the continuum source, and this suffers
from the self-absorption and spatial filtering at the velocity of the
cloud described above.

\subsection{Relationship between the hot core and the continuum sources}

\begin{figure}
$$\includegraphics[scale=0.8]{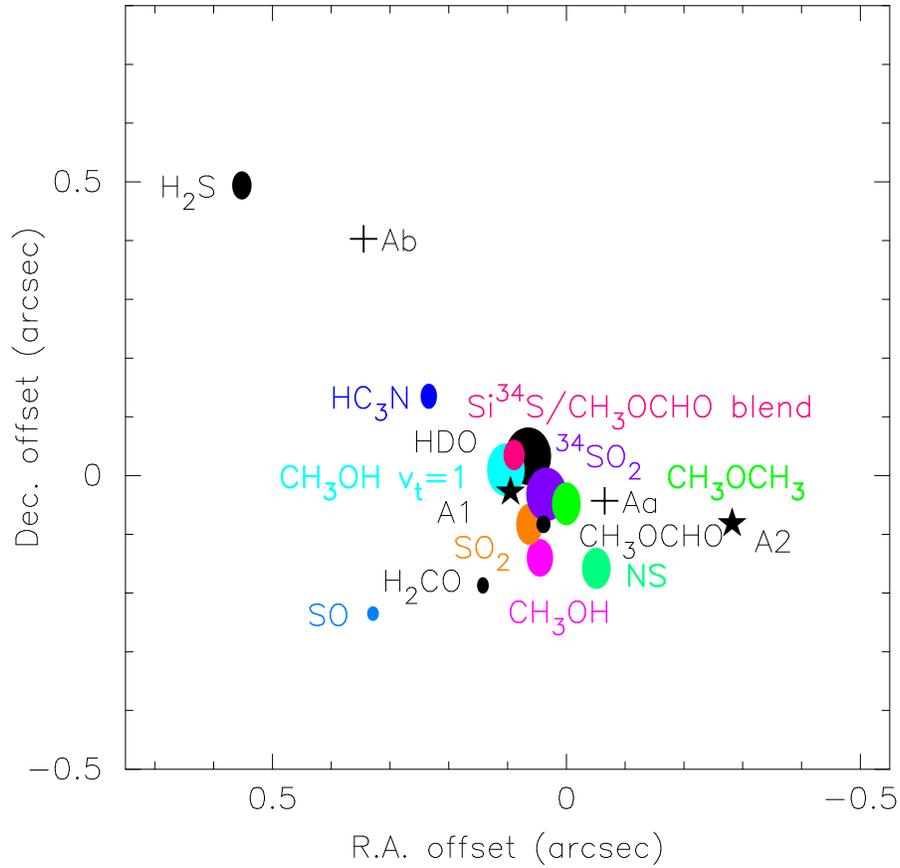}$$
\caption{Location of the integrated emission peaks for the molecular
species detected from IRAS~16293$-$2422 (ellipses) and the continuum
sources Aa, Ab (crosses), A1, and A2 (stars).  The ellipse semi-axes
represent the 1$\sigma$ uncertainties in the positions.  R.A. and dec.\
offsets are relative to the peak in Fig.~\ref{cont_r0}.  \label{hot_core}}
\end{figure}

The emission from species more commonly associated with hot cores in
massive star-forming regions is predominantly located close to source
A\@.  However, the molecular emission is not coincident with either
of the submillimeter continuum peaks.  Fig.~\ref{hot_core} plots the
position of the peak in the integrated emission from each species as
$\pm 1 \sigma$ error ellipses, and shows that all the molecular species
peak to the southeast of a line joining sources Aa and Ab.  A comparison
with Fig.~\ref{overlays} shows that many of the species, but especially
the emission from the high-energy transition of torsionally-excited
methanol, lies in the direction of the shock-ionized radio emission at
source A1, and lends support to our interpretation of A1 as the location
of a clump of very dense gas into which the jet driving the E--W outflow
is ploughing.  Furthermore, the location of the molecular peaks are all
approximately aligned with the directions through which the jet has passed
over the last 20 years.  Thus we infer that the passage of the jet, and
its shock interaction with the dense gas east of Aa, is responsible for
enhancing the abundances of many species at this location, and providing
the high excitation temperatures.

The four molecules that peak somewhat to the south of Aa, i.e., SO,
H$_2$CO, CH$_3$OH, and NS, are all transitions having $E_{\rm upper}
< 100$~K, and the two of these furthest from Aa, SO and H$_2$CO,
have contributions from extended emission.  Most of the high-energy
transitions originate directly east of Aa, close to the shock at A1,
although HC$_3$N and H$_2$S appear at somewhat larger distances along
the direction of the NE--SW outflow.  Note also that those molecules
detected in emission with $T_{\rm rot} \la 90$~K (H$_2$S, H$_2$CO,
SO and CH$_3$OH; Table~\ref{abundances}) lie at some distance from A1.
The remaining molecules (excluding NS, for which $T_{\rm rot}$ has not
been measured) have $T_{\rm rot} \ga 90$~K, and are mostly within 0.1$''$
[$16(D/{\rm 160~pc})$~AU] of A1.  The one exception is HC$_3$N, which
has a high derived $T_{\rm rot}$ and yet lies 0.2$''$ from A1 towards the
northeast.  However, the uncertainty in its $T_{\rm rot}$ is very large,
and is still consistent with being less than 90~K\@.  Further measurements
of intermediate energy transitions are needed to constrain better the
excitation and origin of the HC$_3$N emission from IRAS~16293$-$2422.

\subsection{The unambiguous detection of infall}
\label{infall}

The detection of infall due to gravitational collapse in star forming
regions using molecular emission lines can be confused and complicated
by other motions in the gas, such as rotation and outflow.  The best
examples to date are for sources with simple outflow geometries, for
which the contributions to the kinematics from outflow and rotation can be
quantified or separated (e.g., B335: Zhou et al.\ 1993).  Claims of the
detection of infall for IRAS~16293$-$2422 have been more controversial,
because of the other velocity gradients across the core and the presence
of multiple outflows (Walker et al.\ 1986; Menten et al.\ 1987; Narayanan
et al.\ 1998).  Furthermore, most of the molecular emission line studies
are generally based on data from single-dish telescopes, which trace
the outer envelope on scales of a few $\times 1000$~AU, and not material
collapsing onto the protostar/accretion disk system itself.

The high brightness temperature of the dust continuum emission associated
with sources A and B in IRAS~16293$-$2422 raises the exciting possibility
of using the compact submillimeter emission from the dust disks as a
background source against which to detect redshifted absorption, which
would indicate infall on the size scale of the disk itself.  Ultimately
this is the only unambiguous way of establishing whether gas is infalling
because the absorbing material must be located in front of the protostar,
and any blueshifted emission or absorption must be outflow, while material
in rotation about the central object will have motion perpendicular to the
line of sight, contributing only to absorption at the systemic velocity.
It is important to use a high-dipole moment molecule and a transition that
is likely to have $T_{\rm ex} < T_{\rm B,~dust}$ for such an experiment,
in order to probe the inner parts of the infall envelope.  We have
already shown the molecules such as DNO, present in the cool, outer
envelope of IRAS~16293$-$2422, can be observed in absorption against the
dust emission (Fig.~\ref{deuterated}).  Di Francesco et al.\ (2001) show
that the $3_{1,2}$$-$$2_{1,1}$ transition of formaldehyde at 225.7~GHz is
detected with an inverse P Cygni profile towards the dust emission from
the low-mass protostar NGC~1333~IRAS4, and similar profiles for various
transitions of ammonia and other molecules have been observed against the
free-free continuum emission from HII regions (e.g., Zhang \& Ho 1997;
Zhang, Ho, \& Ohashi 1998).  The best candidate transitions for tracing
the {\it inner} envelope in our data cube are those of formaldehyde and
SO, but the formaldehyde is unfortunately blended with methyl formate
on the redshifted side of the line.  We therefore use SO to investigate
the possibility of detecting collapse in the envelope on size scales of
the background dust emission.  The chemical models of Schoier et al.\
(2002) and Doty et al.\ (2004) suggest high SO abundances only at small
radii, and its excitation temperature also points towards an origin very
close to the embedded protostars.

\begin{figure}
$$\includegraphics[scale=0.65,angle=-90]{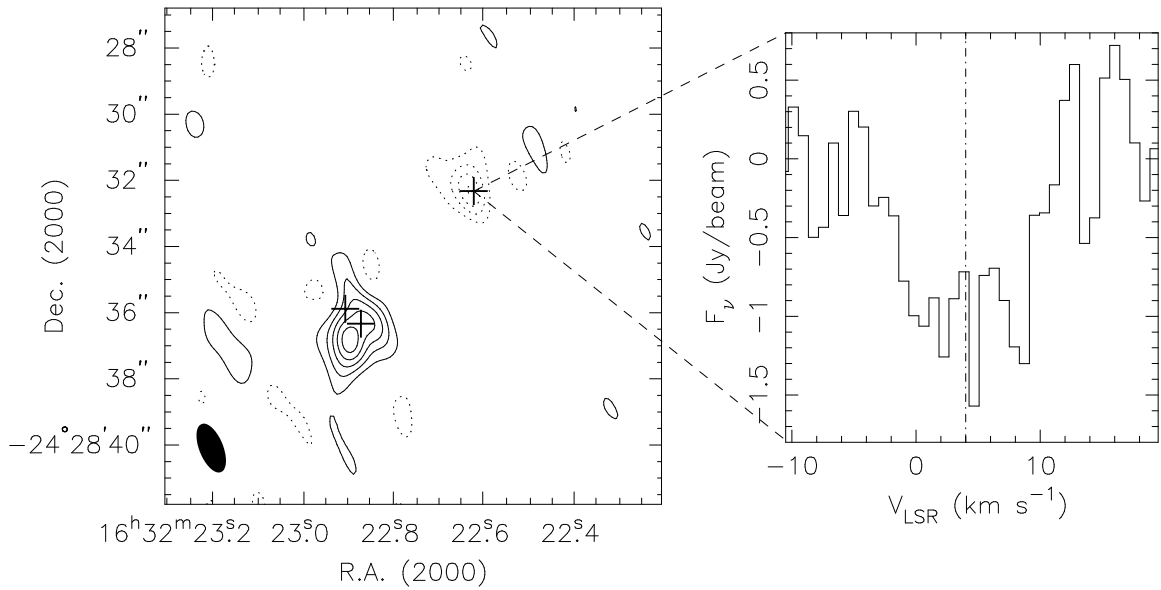}$$
\caption{{\it Left:} image of the integrated SO ($7_7$$-$$6_6$) emission
obtained using an AIPS robust weighting of 0 and excluding baselines
shorter than 55~k$\lambda$.  The resulting CLEAN beam is $1.57'' \times
0.72''$ at P.A. 22, shown in the bottom left corner.  Contours are evenly
spaced at intervals of 3.7~Jy~km~s$^{-1}$.  {\it Right:} spectrum of
the absorption feature at source B, with the systemic $V_{\rm LSR}$
of 4~km~s$^{-1}$ indicated by the dot-dash line.  The rms noise per
channel is 0.5~Jy~beam$^{-1}$.  \label{so_55kl}}
\end{figure}

In Fig.~\ref{so_55kl} we show an image of the SO emission made using
an AIPS robust weighting of 0 and excluding baselines shorter than
55~k$\lambda$, in order to filter out the extended envelope and isolate
compact emission and absorption features.  The emission associated
with the hot core at source A is clear, as is absorption against the
compact dust emission at source B\@.  The spectrum of this absorption
feature is also plotted.  There is significant absorption extending
$\sim 5$~km~s$^{-1}$ both to the redshifted and blueshifted sides of
the systemic velocity of 4~km~s$^{-1}$.  The blueshifted absorption
must be associated with low-velocity outflow from source B (perhaps
its outflow is in the earliest phase of punching its way out of the
surrounding infall envelope), while the redshifted absorption must be due
to envelope material infalling onto the accretion disk.  The rms noise
per channel in the spectrum is 0.5~Jy~beam$^{-1}$, so the significance
of the absorption feature in any one channel is marginal.  However,
reshifted absorption is detected in at least seven consecutive channels,
making this a 5-$\sigma$ detection.  Unfortunately this signal-to-noise
ratio is insufficient to enable detailed modelling of the infall density
and velocity profiles.  However, the infall velocity ($V_{\rm inf} =
\sqrt{2GM/r}$) for a 1~$M_\odot$ central star would suggest that the
material responsible for this absorption is as close as 70~AU to the disk.

\subsection{Conclusions}

We have combined SMA data with archival VLA data to establish the
relationship between the submillimeter and centimeter radio continuum
sources, emission from hot core molecules, and the molecular outflows from
the class 0 protostar IRAS~16293$-$2422.  The overall proper motion of
sources A and B originally noted by Loinard (2002) based on two epochs
of 8~GHz data are now confirmed by other epochs and measurements at
frequencies from 5 to 43~GHz.  Most of this motion is that associated
with the $\rho$ Ophiuchi molecular cloud, but there is also a significant
relative motion of the centimeter radio sources A1 and A2, which excludes
the possibility that these two sources are gravitationally bound unless
A1 is in a highly eccentric orbit and is observed at periastron, the
probability of which is low.

Sources A and B are separated by only 5.2$''$ [$830(D/{\rm 160~pc})$~AU],
yet their properties differ dramatically.  Source B is a single source
at all frequencies observed to date, the highest resolution being 43~GHz
with a beam $86 \times 47$~mas, or $14 \times 7.5$~AU\@.  The continuum
emission originates from optically-thick dust even at a frequency of
8~GHz, and exhibits brightness temperatures of 350--400~K on size scales
of 5--10~AU\@.  Source B is responsible for at least half the total
luminosity of IRAS~16293$-$2422.  The molecular emission from source
B exhibits narrow line widths, with no sign of outflow.  Combined with
its high luminosity (which is presumably dominated by accretion) this
suggests that source B may not even have begun a phase of significant
mass loss yet.

We show that source A, on the other hand, probably comprises three stellar
components.  One is the centimeter radio source A2, which we identify
as the driving source of the large-scale NE--SW molecular outflow.
The second is an as-yet unidentified companion to A2 which drives a
precessing jet responsible for shock-ionized centimeter emission at A1.
The strongest submillimeter continuum emission from source A, denoted
Aa, peaks between the radio sources A1 and A2, but the resolution of
the SMA is insufficient to tell whether this marks the position of
the companion to A2, or the centroid of submillimeter emission from A1
and A2.  We suggest that this companion is responsible for the large-scale
east--west molecular outflow, which is messy and has a wider opening angle
than the NE--SW flow, all of which can be explained by a variable jet
direction.  Finally, we also identify a new submillimeter source 0.64$''$
away from Aa, which we denote Ab.  This source is too far from Aa to be
responsible for the precession of the jet.  Most of the molecular emission
originates from source A, and the overall kinematics of the envelope are
in the opposite sense to that of the outflows, with redshifted emission
to the south, and blueshifted emission to the north.  There is, however,
one compact kinematic feature centered on source A that is similar to
the large-scale east--west outflow, confirming this as its origin.

IRAS~16293$-$2422 is one of only a handful of low-mass protostars which
exhibits emission from molecules more commonly associated with hot cores
in regions of massive star formation.  We show that in this low-mass
source the emission from these molecules is coincident with the shock
at A1.  The abundances derived from the submillimeter emission indicate
significant local enhancements over those expected for a low-mass
protostar for many species, but especially those typically thought
to trace shocks such as the sulphur-bearing molecules SO and SO$_2$.
Our results highlight the key role of shocks in explaining the nature of
hot cores, and demonstrate the importance of high-resolution imaging at
submillimeter wavelengths with the SMA in pinning down rotation diagrams
using high-energy transitions.

Finally, we discuss the possibility of using optically-thick dust emission
from accretion disks as a background source against which to detect
infall in the surrounding envelope.  The advantage of using compact
dust emission instead of the usual technique of using a molecular
emission line as the background source is that the absorption must
be on the physical size scale of the disk itself, and by choosing an
appropriate molecular transition for the absorption it is possible to
isolate the kinematics of infalling gas in the inner envelope, without
confusion from other emission components such as outflow.  When measured
against the dust continuum any blueshifted absorption must be outflow,
rotation will be at the systemic velocity, and redshifted absorption
can be unambiguously identified as infall.  By excluding short baselines
from the SMA data we detect redshifted absorption of SO ($7_7$$-$$6_6$)
against source B extending 5~km~s$^{-1}$ from the systemic $V_{\rm LSR}$.
Unfortunately the signal-to-noise ratio is not sufficient to allow a
detailed analysis of the infall kinematics of IRAS~16293$-$2422, but
clearly this will be a possibility in the future.

\acknowledgments
%\section*{Acknowledgments}

We thank Alison Peck and her colleagues at the SMA for obtaining
the submillimeter data.  We also thank Al Wootten and Mark Claussen
for useful discussions on the nature of IRAS~16293$-$2422, and the
referee for constructive comments that have helped improve the paper.
L.L. acknowledges the financial support of DGAPA, UNAM and CONACyT,
M\'exico.

%\newpage

\newpage

\clearpage

\begin{figure}
$$\includegraphics[scale=0.8]{fig8.ps}$$
\caption{Spectra of the lower and upper sidebands in the SMA data smoothed
over 4 channels (3.25~MHz) for source A (top two panels) and source B
(bottom two panels).  \label{long_spec}}
\end{figure}

\begin{figure}
$$\includegraphics[scale=0.7,angle=-90]{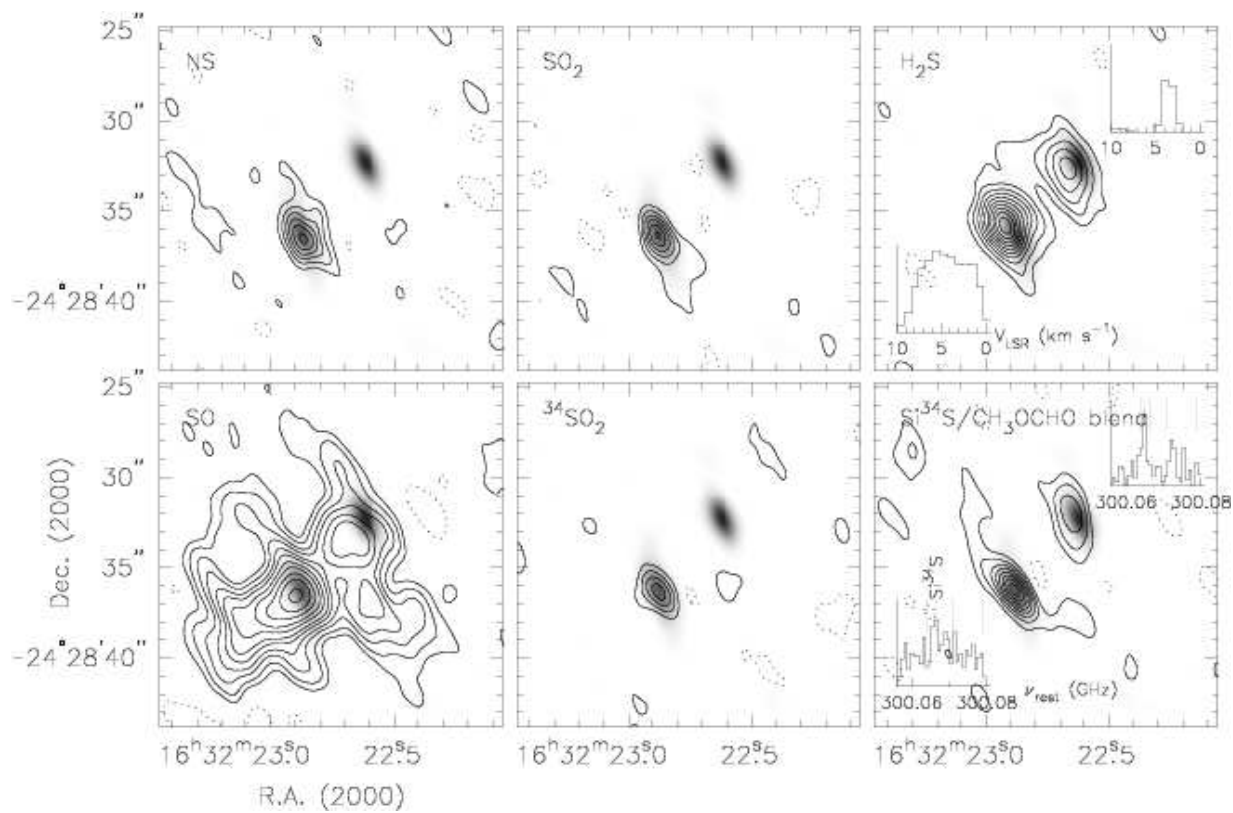}$$
\caption{Emission from sulphur-bearing species.  For display purposes the
H$_2$S emission is integrated over the width of the narrow line associated
with source B, to avoid spectral dilution.  Spectra of sources A and
B are also plotted in the bottom left corner (source A) and top right
corner (source B) of the H$_2$S panel.  In the bottom right panel the
emission is integrated over the line marked in Fig.~\ref{long_spec} as
Si$^{34}$S/CH$_3$OCHO-A/CH$_3$OCHO-E.  Unsmoothed spectra are also plotted
for source A (bottom left corner) and source B (top right corner) in this
panel, with the locations of the lines indicated: the Si$^{34}$S line is
marked explicitly, and the other three lines are transitions of methyl
formate, as listed in Table~\ref{line_ids}.  For all images apart from
SO the contours are linearly spaced at approximately 2$\sigma$ intervals
of 1.43 (NS), 2.00 (SO$_2$), 1.02 (H$_2$S), 1.98 ($^{34}$SO$_2$), and
2.59 (Si$^{34}$S) Jy~km~s$^{-1}$.  For SO the contours are at $-$1,
1, 2, 3.5, 5, 7, 9, 12, 15, 19, 23 and 28 times 2.62~Jy~km~s$^{-1}$.
\label{sulphur}}
\end{figure}

\clearpage

\newpage

\begin{deluxetable}{lccl}
\tablewidth{0pt}
\tablecaption{Summary of VLA continuum archive data.
\label{archive_sum}}
\tablehead{
\colhead{Date} & \colhead{$\nu$ (GHz)} & \colhead{Configuration} &
\colhead{Reference}
}
\startdata
1986 Jul.\ 03 & \phn4.86 & BnA & Wootten (1989) \\
1987 Aug.\ 30 & 14.94 & A & Wootten (1989) \\
1987 Oct.\ 25 & \phn1.49 & BnA & \nodata \\
1987 Nov.\ 01 & \phn1.49 & BnA & \nodata \\
1988 Mar.\ 27 & 14.94 & C & Estalella et al.\ (1991) \\
1988 Mar.\ 27 & 22.46 & C & Estalella et al.\ (1991) \\
1988 Jul.\ 17 & 14.94 & D & Estalella et al.\ (1991) \\
1988 Aug.\ 10 & 14.94 & D & Estalella et al.\ (1991) \\
1988 Aug.\ 20 & 14.94 & D & Estalella et al.\ (1991) \\
1989 Jan.\ 20 & \phn8.44 & A & Mundy et al.\ (1992) \\
1989 Jan.\ 20 & 22.46 & A & Mundy et al.\ (1992) \\
1989 Apr.\ 19 & 14.94 & B & \nodata \\
1992 Apr.\ 30 & \phn8.44 & C & \nodata \\
1992 Apr.\ 30 & 14.94 & C & \nodata \\
1992 Apr.\ 30 & 22.46 & C & \nodata \\
1994 Apr.\ 14 & \phn8.44 & A & Loinard (2002) \\
1994 Apr.\ 15 & \phn8.44 & A & Loinard (2002) \\
1994 Apr.\ 22 & \phn8.44 & A & Loinard (2002) \\
1994 May 20 & \phn8.44 & BnA & Loinard (2002) \\
1998 Jun.\ 16 & 22.25\tablenotemark{a} & BnA & \nodata \\
2002 May 30 & \phn8.46 & BnA & \nodata \\
2003 Jun.\ 22 & 43.34 & A & Rodr\'\i guez et al.\ (2005) \\
2003 Aug.\ 26 & \phn8.46 & A & \nodata \\
\enddata
\tablenotetext{a}{These data were obtained in a 25~MHz filter centered
on 22.25~GHz, with second IF centered on the nearby 22.235~GHz H$_2$O
maser feature.  The maser was then used to correct for tropospheric
phase fluctuations in the continuum data.}
\end{deluxetable}

\begin{deluxetable}{lcc}
\tablewidth{0pt}
\tablecaption{Integrated SMA and VLA continuum flux densities.
\label{int_fluxes}}
\tablehead{
\colhead{$\nu$ (GHz)} & \colhead{$F_\nu$, Source A (mJy)} &
\colhead{$F_\nu$, Source B (mJy)}
}
\startdata
\phn\phn1.49 & $1.32 \pm 0.17$ & $< 0.32~(3\sigma)$ \\
\phn\phn4.86 & $2.81 \pm 0.29$ & $0.281 \pm 0.048$ \\
\phn\phn8.44 & $3.17 \pm 0.76$ & $0.738 \pm 0.089$ \\
\phn14.94\tablenotemark{a} & $4.66 \pm 0.62$ & $2.60 \pm 0.21$ \\
\phn22.46\tablenotemark{b} & $5.00 \pm 0.74$ & $7.28 \pm 1.20$ \\
\phn43.34 & $8.7 \pm 1.6$ & $24.7 \pm 2.5$\phn \\
305 & $3460 \pm 350$\phn & $3150 \pm 320$\phn \\
\enddata
\tablenotetext{a}{The flux density from 1987 August 30 is not included
in the mean quoted here, as discussed in Section~\ref{results_var}.}
\tablenotetext{b}{The mean includes the data obtained at 22.25~GHz.}
\end{deluxetable}

\begin{deluxetable}{lccc}
\tablewidth{0pt}
\tablecaption{Positions and flux densities for compact components.
\label{compact}}
\tablehead{
\colhead{Source} & \colhead{R.A. (J2000)} & \colhead{Dec.\ (J2000)}
& \colhead{$F_{\rm 305~GHz}$ (Jy)}
}
\startdata
Aa & $16^{\rm h} 32^{\rm m} 22\fs8721 \pm 0\fs0010$ & $-24^\circ 28'
36\farcs331 \pm 0\farcs018$ & $\sim 1.6$ \\
Ab & $16^{\rm h} 32^{\rm m} 22\fs9060 \pm 0\fs0028$ & $-24^\circ 28'
35\farcs885 \pm 0\farcs048$ & $\sim 0.5$ \\
B & $16^{\rm h} 32^{\rm m} 22\fs6221 \pm 0\fs0004$ & $-24^\circ 28'
32\farcs326 \pm 0\farcs009$ & $\sim 2.7$ \\
\enddata
\end{deluxetable}

\begin{deluxetable}{lllccr@{$\pm$}l}
\tabletypesize{\small}
\tablewidth{0pt}
\tablecaption{Summary of all lines detected in the lower and upper
sidebands for IRAS~16293$-$2422.\label{line_ids}}
\tablehead{
\colhead{Molecule} & \colhead{Notation} & \colhead{Transition} &
\colhead{Frequency (MHz)} & \colhead{$E_{\rm upper}$ (K)} &
\multicolumn{2}{c}{$\int I_\nu dV$ (Jy~km~s$^{-1}$)\tablenotemark{a}}
}
\startdata
CH$_3$OCH$_3$ & $J_{K_a,K_c}$ & $5_{4,2}$$-$$4_{3,1}$ (AE) & 299884.15 &
                              \phn36.2 & ~~~~~~30.0 & 6.7\tablenotemark{b} \\
	      & & $5_{4,1}$$-$$4_{3,1}$ (EA) & 299886.18 & \phn36.2 &
                                         \multicolumn{2}{c}{\nodata} \\
              & & $5_{4,2}$$-$$4_{3,2}$ (EA) & 299886.79 & \phn36.2 &
                                         \multicolumn{2}{c}{\nodata} \\
              & & $5_{4,1}$$-$$4_{3,2}$ (AE) & 299888.82 & \phn36.2 &
                                         \multicolumn{2}{c}{\nodata} \\
              & & $5_{4,1}$$-$$4_{3,1}$ (EE) & 299889.91 & \phn36.2 &
                                         \multicolumn{2}{c}{\nodata} \\
              & & $5_{4,2}$$-$$4_{3,2}$ (EE) & 299890.95 & \phn36.2 &
                                         \multicolumn{2}{c}{\nodata} \\
              & & $5_{4,2}$$-$$4_{3,1}$ (AA) & 299892.04 & \phn36.2 &
                                         \multicolumn{2}{c}{\nodata} \\
              & & $5_{4,1}$$-$$4_{3,2}$ (AA) & 299896.72 & \phn36.2 &
                                         \multicolumn{2}{c}{\nodata} \\
              & & $8_{3,5}$$-$$7_{2,6}$ (AE) & 299899.47 & \phn45.5 &
                                         15.9 & 4.0\tablenotemark{c} \\
              & & $8_{3,5}$$-$$7_{2,6}$ (EA) & 299899.78 & \phn45.5 &
                                         \multicolumn{2}{c}{\nodata} \\
              & & $8_{3,5}$$-$$7_{2,6}$ (EE) & 299903.00 & \phn45.5 &
                                         \multicolumn{2}{c}{\nodata} \\
              & & $8_{3,5}$$-$$7_{2,6}$ (AA) & 299906.38 & \phn45.5 &
                                         \multicolumn{2}{c}{\nodata} \\
CH$_3$OCHO-A & $J_{K_a,K_c}$ & $24_{6,19}$$-$$23_{6,18}$ & 300063.94 & 202.8 &
                                         \multicolumn{2}{c}{blended} \\
             & & $26_{3,24}$$-$$25_{3,23}$ & 300079.11 & 207.0 &
                                         \multicolumn{2}{c}{blended} \\
             & & $26_{2,24}$$-$$25_{4,22}$ & 300134.83 & 207.0 &
                                         \multicolumn{2}{c}{blended} \\
             & & $27_{1,26}$$-$$26_{2,25}$ & 300834.13 & 210.8 &
                                         \multicolumn{2}{c}{blended} \\
             & & $27_{2,26}$$-$$26_{2,25}$ & 300835.70 & 210.8 &
                                         \multicolumn{2}{c}{blended} \\
             & & $27_{1,26}$$-$$26_{1,25}$ & 300836.94 & 210.8 &
                                         \multicolumn{2}{c}{blended} \\
             & & $27_{2,26}$$-$$26_{1,25}$ & 300838.52 & 210.8 &
                                         \multicolumn{2}{c}{blended} \\
             & & $27_{3,25}$$-$$26_{3,24}$ & 310652.71 & 221.9 &
                                         \multicolumn{2}{c}{blended} \\
             & & $27_{2,25}$$-$$26_{2,24}$ & 310686.06 & 221.9 &
                                         \multicolumn{2}{c}{blended} \\
             & & $26_{3,23}$$-$$25_{3,22}$ & 310693.76 & 216.6 &
                                         \multicolumn{2}{c}{blended} \\
             & & $28_{1,27}$$-$$27_{2,26}$ & 311416.28 & 225.7 &
                                         \multicolumn{2}{c}{blended} \\
             & & $28_{2,27}$$-$$27_{2,26}$ & 311417.16 & 225.7 &
                                         \multicolumn{2}{c}{blended} \\
             & & $28_{1,27}$$-$$27_{1,26}$ & 311417.86 & 225.7 &
                                         \multicolumn{2}{c}{blended} \\
             & & $28_{2,27}$$-$$27_{1,26}$ & 311418.74 & 225.7 &
                                         \multicolumn{2}{c}{blended} \\
             & & $25_{8,18}$$-$$24_{8,17}$ & 311641.97 & 235.5 &
                                         \multicolumn{2}{c}{blended} \\
CH$_3$OCHO-E & $J_{K_a,K_c}$ & $26_{5,22}$$-$$25_{5,21}$ & 300070.76 & 207.0 &
                                         \multicolumn{2}{c}{blended} \\
             & & $26_{4,23}$$-$$25_{4,22}$ & 300126.62 & 207.0 &
                                         \multicolumn{2}{c}{blended} \\
             & & $36_{7,30}$$-$$36_{5,32}$ & 300399.6\phn & 393.5 &
                                         \multicolumn{2}{c}{blended} \\
             & & $36_{7,30}$$-$$36_{6,31}$ & 300399.9\phn & 393.5 &
                                         \multicolumn{2}{c}{blended} \\
             & & $25_{5,20}$$-$$24_{5,19}$ & 300403.20 & 201.7 &
                                         \multicolumn{2}{c}{blended} \\
             & & $36_{6,30}$$-$$36_{5,32}$ & 300408.1\phn & 393.5 &
                                         \multicolumn{2}{c}{blended} \\
             & & $36_{6,30}$$-$$36_{6,31}$ & 300408.4\phn & 393.5 &
                                         \multicolumn{2}{c}{blended} \\
             & & $27_{3,25}$$-$$26_{2,24}$ & 300829.09 & 210.8 &
                                         \multicolumn{2}{c}{blended} \\
             & & $27_{2,25}$$-$$26_{2,24}$ & 300830.68 & 210.8 &
                                         \multicolumn{2}{c}{blended} \\
             & & $27_{3,25}$$-$$26_{3,24}$ & 300831.93 & 210.8 &
                                         \multicolumn{2}{c}{blended} \\
             & & $27_{2,25}$$-$$26_{3,24}$ & 300833.51 & 210.8 &
                                         \multicolumn{2}{c}{blended} \\
             & & $27_{5,23}$$-$$26_{5,22}$ & 310644.51 & 221.9 &
                                         \multicolumn{2}{c}{blended} \\
             & & $27_{4,24}$$-$$26_{4,23}$ & 310677.95 & 221.9 &
                                         \multicolumn{2}{c}{blended} \\
             & & $26_{5,21}$$-$$25_{5,20}$ & 310684.19 & 216.6 &
                                         \multicolumn{2}{c}{blended} \\
             & & $15_{5,10}$$-$$14_{4,11}$ & 311408.99 & \phn87.9 &
                                         \multicolumn{2}{c}{blended} \\
             & & $28_{3,26}$$-$$27_{2,25}$ & 311411.35 & 225.7 &
                                         \multicolumn{2}{c}{blended} \\
             & & $28_{2,26}$$-$$27_{2,25}$ & 311412.1\phn & 225.7 &
                                         \multicolumn{2}{c}{blended} \\
             & & $28_{3,26}$$-$$27_{3,25}$ & 311412.7\phn & 225.7 & 
                                         \multicolumn{2}{c}{blended} \\
             & & $28_{2,26}$$-$$27_{3,25}$ & 311413.81 & 225.7 &
                                         \multicolumn{2}{c}{blended} \\
             & & $25_{8,18}$$-$$24_{8,17}$ & 311637.10 & 235.5 &
                                         \multicolumn{2}{c}{blended} \\
Si$^{34}$S (?) & $J$ & 17$-$16 & 300066\phd\phn\phn & 129.7 &
                                         \multicolumn{2}{c}{blended} \\
NS ($^2\Pi_{1/2}$) & $J_F$ & $6.5_{15/2}$$-$$5.5_{13/2}\, e$ & 300097.10 &
                              \phn54.4 & ~~~~~~12.2 & 2.8\tablenotemark{d} \\
                   & & $6.5_{13/2}$$-$$5.5_{11/2}\, e$ & 300098.61 & \phn54.4 &
                                         \multicolumn{2}{c}{\nodata} \\
                   & & $6.5_{11/2}$$-$$5.5_{9/2}\, e$ & 300098.61 & \phn54.4 &
                                         \multicolumn{2}{c}{\nodata} \\
HC$_3$N & $J$ & 33$-$32 & 300159.65 & 245.1 & 30.5 & 6.4 \\
SO$_2$ & $J_{K_a,K_c}$ & $32_{3,29}$$-$$32_{2,30}$ & 300273.42 &
                                                  519.1 & 14.6 & 3.5 \\
H$_2$S & $J_{K_a,K_c}$ & $3_{3,0}$$-$$3_{2,1}$ & 300505.56 & 169.0 & 77 & 17 \\
CH$_3$OH $v_t=1$ & $J_{K_a,K_c}$ & $16_{3,14}E$$-$$17_{4,14}E$ &
                                    300763.4\phn & 732.4 & 4.0 & 1.1 \\
H$_2$CO & $J_{K_a,K_c}$ & $4_{1,3}$$-$$3_{1,2}$ & 300836.64 & \phn47.9 &
                                           397 & 80\tablenotemark{e} \\
DNO & $J_{(K_a,K_c)_F}$\tablenotemark{f} & $4_{(1,3)_3}$$-$$3_{(1,2)_3}$ &
   300952.3\phn & \phn49.5 & \multicolumn{2}{c}{absorption\tablenotemark{g}} \\
    & & $4_{(1,3)_5}$$-$$3_{(1,2)_4}$ & 300955.14 & \phn49.5 &
                                         \multicolumn{2}{c}{\nodata} \\
    & & $4_{(1,3)_4}$$-$$3_{(1,2)_3}$ & 300955.23 & \phn49.5 &
                                         \multicolumn{2}{c}{\nodata} \\
    & & $4_{(1,3)_3}$$-$$3_{(1,2)_2}$ & 300955.33 & \phn49.5 &
                                         \multicolumn{2}{c}{\nodata} \\
    & & $4_{(1,3)_4}$$-$$3_{(1,2)_4}$ & 300957.5\phn & \phn49.5 &
                                         \multicolumn{2}{c}{\nodata} \\
SO & $N_J$ & $7_7$$-$$6_6$ & 301286.12 & \phn71.0 & 582 & 117 \\
CH$_3$OH & $J_{K_a,K_c}$ & $3_{1,3}E$$-$$2_{0,3}E$ & 310193.02 &
                                               \phn35.0 & 27.7 & 6.1 \\
HDO & $J_{K_a,K_c}$ & $5_{2,3}$$-$$5_{2,4}$ & 310533.29 & 437.7 & 12.0 & 3.0 \\
$^{34}$SO$_2$ & $J_{K_a,K_c}$ & $17_{1,17}$$-$$16_{0,16}$ &
                      311485.38 & 135.6 & 7.6 & 1.9\tablenotemark{h} \\
              & & $20_{4,16}$$-$$20_{3,17}$ & 311487.37 & 231.4 &
                                         \multicolumn{2}{c}{\nodata} \\
Unidentified\tablenotemark{i} & \nodata & \nodata & 300889\phd\phn\phn & \nodata &
                                         \multicolumn{2}{c}{\nodata} \\
Unidentified & \nodata & \nodata & 301077\phd\phn\phn & \nodata &
                                         \multicolumn{2}{c}{\nodata} \\
Unidentified & \nodata & \nodata & 310370\phd\phn\phn & \nodata &
                                         \multicolumn{2}{c}{\nodata} \\
Unidentified & \nodata & \nodata & 310505\phd\phn\phn & \nodata &
                                         \multicolumn{2}{c}{\nodata} \\
Unidentified & \nodata & \nodata & 310568\phd\phn\phn & \nodata &
                                         \multicolumn{2}{c}{\nodata} \\
Unidentified & \nodata & \nodata & 311023\phd\phn\phn & \nodata &
                                         \multicolumn{2}{c}{\nodata} \\
Unidentified & \nodata & \nodata & 311208\phd\phn\phn & \nodata &
                                         \multicolumn{2}{c}{\nodata} \\
Unidentified & \nodata & \nodata & 311313\phd\phn\phn & \nodata &
                                         \multicolumn{2}{c}{\nodata} \\
\enddata
\tablenotetext{a}{The emission is integrated over an area enclosing
the 2-$\sigma$ contour, and includes emission from both A and B where
appropriate, for comparison with single-dish measurements.  The error
quoted includes both statistical and calibration uncertainties.}
\tablenotetext{b}{Integrated flux is for a blend of all 5$-$4
transitions.}
\tablenotetext{c}{Integrated flux is for a blend of all 8$-$7
transitions.}
\tablenotetext{d}{Integrated flux is for a blend of all 6.5$-$5.5
transitions.}
\tablenotetext{e}{Includes several blended lines of methyl formate,
which are estimated to contribute 5--10\% of the total integrated flux.}
\tablenotetext{f}{F quantum number refers to N hyperfine splitting.}
\tablenotetext{g}{Absorption feature is a blend of all the 4$-$3
transitions.}
\tablenotetext{h}{Integrated flux is for a blend of the
$17_{1,17}$$-$$16_{0,16}$ and $20_{4,16}$$-$$20_{3,17}$ transitions.}
\tablenotetext{i}{Observed in emission toward source A\@.  The absorption
toward source B on the redshifted side of this feature may be the same
line and so has not been listed separately.}
\end{deluxetable}

\begin{deluxetable}{lccr@{$\times$}lr@{$\times$}lr@{$\times$}lr@{$\times$}l}
\tablewidth{0pt}
\tabletypesize{\small}
\tablecaption{Rotation temperatures, emission sizes, column densities,
and abundances for the molecules detected.
\label{abundances}}
\tablehead{
\colhead{Molecule} & \colhead{T$_{\rm rot}$} & \colhead{Emission area} &
\multicolumn{2}{c}{$N_{\rm mol}$} &
\multicolumn{2}{c}{$N_{\rm mol}/N_{\rm H_2}$\tablenotemark{a}} &
\multicolumn{2}{c}{$f_{\rm in}(X)$\tablenotemark{b}} &
\multicolumn{2}{c}{$X_{\rm max}$\tablenotemark{c}} \\
\colhead{} & \colhead{(K)} & \colhead{(arcsec$^2$)} &
\multicolumn{2}{c}{(cm$^{-2}$)} & \colhead{} & \colhead{} &
\colhead{}}
\startdata
CH$_3$OCH$_3$ & 95$\pm$73 & \phn20.7 & 6.4 & $10^{15}$ & 7.6 & $10^{-8}$ &
  $<4.0$ & $10^{-8}$ & 3  & $10^{-9}$ \\
HC$_3$N & 320$\pm$300 & \phn11.2 & 1.6 & $10^{14}$ & 1.0 & $10^{-9}$ &
  1.0 & $10^{-9}$ & 3  &$10^{-10}$ \\
CH$_3$OH & \phantom{\tablenotemark{d}}\phn85\tablenotemark{d}\phn & \phn10.5 &
  1.7 & $10^{16}$ & 9.4 & $10^{-8}$ & 3.0 & $10^{-7}$ & 1 & $10^{-7}$ \\
H$_2$CO & \phantom{\tablenotemark{d}}\phn80\tablenotemark{d}\phn & 118.5 &
  3.2 & $10^{14}$ & 1.1 & $10^{-7}$ & 6.0 & $10^{-8}$ & 6 & $10^{-8}$ \\
NS & \phantom{\tablenotemark{e}}100\tablenotemark{e}\phn & \phn\phn7.2 & 2.3 &
  $10^{14}$ & 6.4 & $10^{-10}$ & \multicolumn{2}{c}{\nodata} &
  \multicolumn{2}{c}{\nodata} \\
H$_2$S & 60$\pm$7\phn & \phn37.0 & 2.9 & $10^{15}$ & 9.4 & $10^{-8}$ &
  9.0 & $10^{-8}$ & 3  & $10^{-9}$ \\
SO & \phantom{\tablenotemark{f}}\phn80\tablenotemark{f}\phn & 132.2 &
  1.2 & $10^{15}$ & 5.0 & $10^{-7}$ & 2.5 & $10^{-7}$ & 4  & $10^{-9}$ \\
SO$_2$ & \phantom{\tablenotemark{g}}135$\pm$20\tablenotemark{g}\phn &
  \phn\phn7.7 & 2.9 & $10^{16}$ & 9.7 & $10^{-8}$ & 1.0 & $10^{-7}$ &
  1 & $10^{-8}$ \\
$^{34}$SO$_2$ & \phantom{\tablenotemark{h}}135\tablenotemark{h}\phn &
  \phn\phn3.3\tablenotemark{i} & 1.7 & $10^{15}$ & 1.3 & $10^{-9}$ &
  \multicolumn{2}{c}{\nodata} & \multicolumn{2}{c}{\nodata} \\
DNO & $\sim$15--20\phantom{$\sim$} & \nodata\tablenotemark{j} & $\sim 3$ &
  $10^{15}$ & $\sim 2$ & $10^{-9}$ & \multicolumn{2}{c}{\nodata} &
  \multicolumn{2}{c}{\nodata} \\
HDO & 236$\pm$44\phn & \phn\phn3.4 & 2.9 & $10^{16}$ & 2.2 & $10^{-8}$ &
  \multicolumn{2}{c}{\nodata} & \multicolumn{2}{c}{\nodata} \\
\enddata
\tablenotetext{a}{Mean $N_{\rm H_2}$ over the area of the emission for
each molecule from the current work, as described in the text.}
\tablenotetext{b}{Abundances from Table~7 of Sch\"oier et al.\ (2002),
which assume an abundance ``jump'' inside the radius where the temperature
is $>90$~K (i.e., $r<2\times 10^{15}$~cm).}
\tablenotetext{c}{Approximate maximum abundance from the physical-chemical
model plots of Doty et al.\ (2004), for $r<2.5\times 10^{15}$~cm.}
\tablenotetext{d}{From van Dishoeck et al.\ (1995).}
\tablenotetext{e}{Assumed value.}
\tablenotetext{f}{Derived from $^{34}$SO by Blake et al.\ (1994).}
\tablenotetext{g}{Derived using lines with $E_{\rm upper} > 200$~K
only.}
\tablenotetext{h}{From the SO$_2$ rotation diagram analysis.}
\tablenotetext{i}{Emission is unresolved, so the emission area is
assumed to be that of the synthesized beam.}
\tablenotetext{j}{Absorption against source A.}
\end{deluxetable}


\begin{references}

\reference{}Anglada, G. 1995, Rev.\ Mex.\ Astr.\ Ap.\ (Ser.\ de
Conf.), 1, 67
\reference{}Baars, J. W. M., Genzel, R., Pauliny-Toth, I. I. K., \&
Witzel, A. 1977, A\&A, 61, 99
\reference{}Bally, J., \& Devine, D. 1994, ApJ, 428, L65
\reference{}Bate, M. R., Bonnell, I. A., Clarke, C. J., Lubow, S. H.,
Ogilvie, G. I., Pringle, J. E., \& Tout, C. A. 2000, MNRAS, 317, 773
\reference{}Bence, S. J., Richer, J. S., \& Padman, R. 1996, MNRAS,
279, 866
\reference{}Blake, G. A., van Dishoeck, E. F., Jansen, D. J.,
Groesbeck, T. D., \& Mundy, L. G. 1994, ApJ, 428, 680
\reference{}Bottinelli, S., Ceccarelli, C., Neri, R., Williams, J.
P., Caux, E., Cazaux, S., Lefloch, B., Maret, S., \& Tielens, A. G.
G. M. 2004, ApJ, 617, L69
\reference{}Briggs, D. S. 1995, PhD Thesis, New Mexico Institute of
Mining and Technology
\reference{}Cabrit, S., \& Bertout, C. 1992, A\&A, 261, 274
\reference{}Caselli, P., Hasegawa, T. I., \& Herbst, E. 1993, ApJ,
408, 548
\reference{}Cazaux, S., Tielens, A. G. G. M., Ceccarelli, C.,
Castets, A., Wakelam, V., Caux, E., Parise, B., \& Teyssier, D.
2003, ApJ, 593, L51
\reference{}Ceccarelli, C., Castets, A., Loinard, L., Caux, E., \&
Tielens, A. G. G. M. 1998, A\&A, 338, L43
\reference{}Ceccarelli, C., Loinard, L., Castets, A., Tielens, A. G.
G. M., \& Caux, E. 2000, A\&A, 357, L9
\reference{}Ceccarelli, C., Loinard, L., Castets, A., Tielens, A. G.
G. M., Caux, E., Lefloch, B., \& Vastel, C. 2001, A\&A, 372, 998
%\reference{}Charnley, S. B. 1997, ApJ, 481, 396
\reference{}Charnley, S. B., Tielens, A. G. G. M., \& Millar, T. J.
1992, ApJ, 399, L71
\reference{}Chini, R. 1981, A\&A, 99, 346
\reference{}Curiel, S., Cant\'o, J., \& Rodr\'\i guez, L. F. 1987,
Rev.\ Mex.\ Astr.\ Ap., 14, 595
\reference{}Curiel, S., Girart, J. M., Rodr\'\i guez, L. F., \&
Cant\'o, J. 2003, ApJ, 582, L109
\reference{}Di Francesco, J., Myers, P. C., Wilner, D. J., Ohashi,
N., \& Mardones, D. 2001, ApJ, 562, 770
\reference{}Doty, S. D., Sch\"oier, F. L., \& van Dishoeck, E. F.
2004, A\&A, 418, 1021
\reference{}Estalella, R., Anglada, G., Rodr\'\i guez, L. F., \&
Garay, G. 1991, ApJ, 371, 626
\reference{}Garay, G., Ram\'\i rez, S., Rodr\'\i guez, L. F.,
Curiel, S., \& Torrelles, J. M. 1996, ApJ, 459, 193
\reference{}Ghavamian, P., \& Hartigan, P. 1998, ApJ, 501, 687
\reference{}Goldsmith, P. F., \& Langer, W. D. 1999, ApJ, 517, 209
\reference{}Gonz\'alez, R. F., \& Cant\'o, J. 2002, ApJ, 580, 459
\reference{}Gueth, F., Guilloteau, S., Dutrey, A., \& Bachiller, R.
1997, A\&A, 323, 943
\reference{}Hatchell, J., Thompson, M. A., Millar, T. J., \&
Macdonald, G. H. 1998, A\&A, 338, 713
\reference{}Hildebrand, R. H. 1983, QJRAS, 24, 267
\reference{}Hirano, N., Mikami, H., Umemoto, T., Yamamoto, S., \&
Taniguchi, T. 2001, ApJ, 547, 899
\reference{}Horn, A., M\o llendal, H., Sekiguchi, O., Uggerud, E.,
Roberts, H., Herbst, E., Viggiano, A. A., \& Fridgen, T. D. 2004,
ApJ, 611, 605
\reference{}Knude, J., \& H\o g, E. 1998, A\&A, 338, 897
\reference{}Kuan, Y.-J., et al.\ 2004, ApJ, 616, L27
\reference{}Loinard, L. 2002, Rev.\ Mex.\ Astr.\ Ap., 38, 61
\reference{}Loinard, L., Castets, A., Ceccarelli, C., Tielens, A. G.
G. M., Faure, A., Caux, E., \& Duvert, G. 2000, A\&A, 359, 1169
\reference{}Looney, L. W., Mundy, L. G., \& Welch, W. J. 2000, ApJ,
529, 477
\reference{}Menten, K. M., Serabyn, E., G\"usten, R., \& Wilson, T.
L. 1987, A\&A, 177, L57
\reference{}Millar, T. J., Farquhar, P. R. A., \& Willacy, K. 1997,
A\&AS, 121, 139
\reference{}Millar, T. J., Herbst, E., \& Charnley, S. B. 1991, ApJ,
369, 147
\reference{}Minh, Y. C., Ziurys, L. M., Irvine, W. M., \& McGonagle,
D. 1990, ApJ, 360, 136
\reference{}Miyake, K., \& Nakagawa, Y. 1993, Icarus, 106, 20
\reference{}Mizuno, A., Fukui, Y., Iwata, T., Nozawa, S., \& Takano,
T. 1990, ApJ, 356, 184
\reference{}M\"uller, H. S. P., Thorwirth, S., Roth, D. A., \&
Winnewisser, G. 2001, A\&A, 370, L49
\reference{}Mundy, L. G., Wilking, B. A., \& Myers, S. T. 1986, ApJ,
311, L75
\reference{}Mundy, L. G., Wootten, H. A., Wilking, B. A., Blake, G.
A., \& Sargent, A. I. 1992, ApJ, 385, 306
\reference{}Narayanan, G., Walker, C. K., \& Buckley, H. D. 1998,
ApJ, 496, 292
\reference{}Neufeld, D. A., \& Hollenbach, D. J. 1996, ApJ, 471, L45
\reference{}Ossenkopf, V., \& Henning, Th. 1994, A\&A, 291, 943
\reference{}Parise, B., Castets, A., Herbst, E., Caux, E.,
Ceccarelli, C., Mukhopadhyay, I., \& Tielens, A. G. G. M. 2004a,
A\&A, 416, 159
\reference{}Parise, B., et al.\ 2004b, A\&A, in press
\reference{}Patience, J., Ghez, A. M., Reid, I. N., \& Matthews, K.
2002, AJ, 123, 1570
\reference{}Pearson, T. J., \& Readhead, A. C. S. 1984, ARA\&A, 22, 97
\reference{}Pickett, H. M., Poynter, R. L., Cohen, E. A., Delitsky,
M. L., Pearson, J. C., \& M\"uller, H. S. P. 1998, J. Quant.\
Spectrosc.\ \& Rad.\ Transfer, 60, 883
\reference{}Preibisch, T., Sonnhalter, C., \& Yorke, H. W. 1995,
A\&A, 299, 144
\reference{}Reynolds, S. P. 1986, ApJ, 304, 713
\reference{}Richer, J. S., \& Padman, R. 1991, MNRAS, 251, 707
\reference{}Roberts, H., Fuller, G. A., Millar, T. J., Hatchell, J.,
\& Buckle, J. V. 2002, A\&A, 381, 1026
\reference{}Rodr\'\i guez, L. F., Ho, P. T. P., Torrelles, J. M.,
Curiel, S., \& Cant\'o, J. 1990, ApJ, 352, 645
\reference{}Rodr\'\i guez, L. F., Loinard, L., D'Alessio, P.,
Wilner, D. J., \& Ho, P. T. P. 2005, ApJ, in press
\reference{}Sch\"oier, F. L., J\o rgensen, J. K., van Dishoeck, E.
F., \& Blake, G. A. 2002, A\&A, 390, 1001
\reference{}Sch\"oier, F. L., J\o rgensen, J. K., van Dishoeck, E.
F., \& Blake, G. A. 2004, A\&A, 418, 185
\reference{}Schwartz, R. D., \& Greene, T. P. 1999, AJ, 117, 456
\reference{}Serabyn, E., \& Weisstein, E. W. 1995, ApJ, 451, 238
\reference{}Shang, H., Lizano, S., Glassgold, A., \& Shu, F. 2004,
ApJ, 612, L69
\reference{}Snyder, L. E., Kuan, Y.-J., Ziurys, L. M., \& Hollis, J.
M. 1993, ApJ, 403, L17
\reference{}Stark, R., et al.\ 2004, ApJ, 608, 341
\reference{}Suzuki, H., Yamamoto, S., Ohishi, M., Kaifu, N.,
Ishikawa, S.-I., Hirahara, Y., \& Takano, S. 1992, ApJ, 392, 551
\reference{}Terebey, S., Shu, F. H., \& Cassen, P. C. 1984, ApJ,
286, 529
\reference{}Terquem, C., Eisl\"offel, J., Papaloizou, J. C. B., \&
Nelson, R. P. 1999, ApJ, 512, L131
\reference{}Tohline, J. E. 2002, ARA\&A, 40, 349
\reference{}van der Tak, F. F. S., Boonman, A. M. S., Braakman, R.,
\& van Dishoeck, E. F. 2003, A\&A, 412, 133
\reference{}van Dishoeck, E. F., \& Blake, G. A. 1998, ARA\&A, 36, 317
\reference{}van Dishoeck, E. F., Blake, G. A., Jansen, D. J., \&
Groesbeck, T. D. 1995, ApJ, 447, 760
\reference{}Viti, S., Caselli, P., Hartquist, T. W., \& Williams, D.
A. 2001, A\&A, 370, 1017
\reference{}Wakelam, V., Caselli, P., Ceccarelli, C., Herbst, E., \&
Castets, A. 2004a, A\&A, 422, 159
\reference{}Wakelam, V., Castets, A., Ceccarelli, C., Lefloch, B.,
Caux, E., \& Pagani, L. 2004b, A\&A, 413, 609
\reference{}Walker, C. K., Lada, C. J., Young, E. T., Maloney, P.
R., \& Wilking, B. A. 1986, ApJ, 309, L47
\reference{}Walker, C. K., Carlstrom, J. E., \& Bieging, J. H. 1993,
ApJ, 402, 655
\reference{}Wilson, T. L., \& Rood, R. T. 1994, ARA\&A, 32, 191
%\reference{}Wolfire, M. G., \& Churchwell, E. 1994, ApJ, 427, 889
\reference{}Wootten, A. 1989, ApJ, 337, 858
\reference{}Wootten, A., Claussen, M., Marvel, K., \& Wilking, B.
1999, in ``The Physics and Chemistry of the Interstellar Medium,'' eds V.
Ossenkopf, J. Stutzki, \& G. Winnewisser (GCA-Verlag, Herdecke), 295
\reference{}Zhang, Q., \& Ho, P. T. P. 1997, ApJ, 488, 241
\reference{}Zhang, Q., Ho, P. T. P., \& Ohashi, N. 1998, ApJ, 494, 636
\reference{}Zhou, S., Evans, N. J., K\"ompe, C., \& Walmsley, C. M.
1993, ApJ, 404, 232
\reference{}Zhou, S. 1995, ApJ, 442, 685
\reference{}Ziurys, L. M., Hollis, J. M., \& Snyder, L. E. 1994,
ApJ, 430, 706

\end{references}
\end{document}